\documentclass[]{aastex631}

\usepackage{amsmath,amssymb}
\usepackage{multirow}

\shorttitle{tSZ and kSZ effects fitting formula}
\shortauthors{K. K. Lee et al.}
\graphicspath{{./}{figures/}}

\begin{document}

\title{An exploration of the properties of cluster profiles for the thermal and kinetic Sunyaev-Zel'dovich effects}

\author[0000-0001-6255-2773]{Billy K. K. Lee}
\affiliation{Department of Physics and Institute of Theoretical Physics, The Chinese University of Hong Kong, Shatin, N.T., Hong Kong SAR, People's Republic of China}

\author[0000-0002-1297-3673]{William R. Coulton}
\affiliation{Center for Computational Astrophysics, Flatiron Institute, New York, NY 10010, USA}

\author[0000-0003-2911-9163]{Leander Thiele}
\affiliation{Princeton University, Princeton, NJ 08540, USA}

\author{Shirley Ho}
\affiliation{Center for Computational Astrophysics, Flatiron Institute, New York, NY 10010, USA}
\affiliation{Princeton University, Princeton, NJ 08540, USA}
\affiliation{New York University, New York, NY 10010, USA}
\affiliation{Carnegie Mellon University, Pittsburgh, PA 15289, USA}



\begin{abstract}
With the advent of high-resolution, low-noise CMB measurements, the ability to extract cosmological information from thermal Sunyaev-Zel’dovich effect and kinetic Sunyaev-Zel’dovich effect will be limited not by statistical uncertainties but rather by systematic and theoretical uncertainties. The theoretical uncertainty is driven by the lack of knowledge about the electron pressure and density. Thus we explore the electron pressure and density distributions in the IllustrisTNG hydrodynamical simulations, and we demonstrate that the cluster properties exhibit a strong dependence on the halo concentration -- providing some of the first evidence of cluster assembly bias in the electron pressure and density. Further, our work shows evidence for a broken power-law mass dependence, with lower pressure in lower mass halos than previous work and a strong evolution with mass of the radial correlations in the electron density and pressure. Both of these effects highlight the differing impact of active galactic nuclei and supernova feedback on the gas in galaxy groups compared to massive clusters. We verified that we see qualitatively similar features in the SIMBA hydro-dynamical simulations, suggesting these effects could be generic features. Finally, we provide a parametric formula for the electron pressure and density profile as a function of dark matter halo mass, halo concentration, and redshift. These fitting formulae can reproduce the distribution of density and pressure of clusters and will be useful in extracting cosmological information from upcoming CMB surveys. 

\end{abstract}

\keywords{}


\section{Introduction} \label{sec:intro}

The Cosmic Microwave Background (CMB) is a powerful probe of processes throughout the history of the universe. Through studying the primary CMB anisotropies, we have learnt about the geometry and age of the universe, its energy content, and the properties of the primordial fluctuations \citep{1965ApJ...142..419P,1965ApJ...142..414D,Hinshaw_2013, Planck:2015fie}. By studying CMB secondary anisotropies, anisotropies generated by processes between the observer and surface of last scattering, we can learn about the cosmic star formation rate \citep{Planck_XXX_2013}, the properties of galaxy clusters and the distribution of matter, as well as fundamental physics, such as the neutrino mass sum \citep{RoyChoudhury:2018gay, Vagnozzi:2018jhn}. CMB secondaries are sourced by a range of processes, and this work focuses on those arising from scattering between CMB photons with free electrons, the thermal and kinetic Sunyaev–Zel'dovich (tSZ and kSZ) effects \citep{Zeldovich:1969ff, Sunyaev:1970eu, 1972CoASP...4..173S, 1980ARA&A..18..537S}. The tSZ effect is caused by the scattering of CMB photons of hot electrons in galaxy clusters, whereas the kSZ effect occurs as CMB photons are scattered off electrons with coherent velocities along the line of sight.

The tSZ effect is of particular interest as different aspects of its signature encode different information. First, the tSZ effect is a powerful method to detect massive clusters, especially as the ability to detect clusters of a given mass is largely independent of redshift \citep{Planck:2013shx, Hasselfield:2013wf, Reichardt:2012yj}. The integrated tSZ signal, known as the integrated Compton-$y$ parameter, for individual clusters can be used as a mass proxy through the use of scaling relations \citep{Salvati:2019zdp, Czakon:2014wga} and the resulting calibrated tSZ cluster counts, particularly when combined with measurements of their redshifts, are a very powerful probe of cosmology. For example, the counts are highly sensitive to the sum of the neutrino masses and the dark energy equation of state \citep{Makiya:2019lvm, Horowitz:2016dwk, Planck:2015vgm, Hurier:2017jgi, Salvati:2017rsn,Madhavacheril_2017}. 

A second approach to using the tSZ effect is to probe the spatial distribution of the Compton-$y$ signal through $N$-point functions. Low-mass halos that are too faint to detect individually can contribute to the tSZ power spectrum significantly \citep{Komatsu_2002,Planck:2015vgm, 2010ApJ...725...91B}. On the other hand, the bispectrum is more sensitive to the massive halos at intermediate redshifts \citep{2012ApJ...760....5B,Crawford_2014,Coulton_2018}, where X-ray imaging is available \citep{Arnaud:2009tt}. As a result, a combination of summary statistics that weigh halos differently offers us a tool to study a wide mass range of halos. The strong astrophysical dependence on the spatial properties of the tSZ effect encodes the details of the circumgalactic medium and the intracluster medium \citep{Kim:2021gtf, Moser:2021llm}, including the hydrodynamics of the gas, star-formation, and active galactic nuclei (AGN) activities \citep{Battaglia:2017neq, 2018ApJ...865..109S, Tanimura:2021kyo}. There are existing measurements of the tSZ power spectrum by Planck \citep{Planck:2015vgm}, Atacama Cosmology Telescope (ACT) \citep{ACT:2020frw}, and the South Pole Telescope (SPT) \citep{Shirokoff:2010cs}. These measurements can be used to probe the amplitude of density fluctuations  \citep[e.g.][]{Planck:2015vgm}. The precision of these measurements will increase significantly in upcoming Simons Observatory and CMB-S4 measurements \citep{SimonsObservatory:2018koc,CMBS4}. 

The kSZ effect is likewise a powerful and versatile astrophysical probe. The kSZ effect is sensitive to both the gas velocity and the gas density, and these two aspects allow us to probe a wide range of phenomena. First, focusing on the former, through pairwise \citep{Hand_2012} and bispectrum estimators \citep{Smith_2018}, it is possible to isolate the large-scale velocity field through kSZ measurements. This field is a rich source of information on primordial non-Gaussianity \citep{Munchmeyer_2019}, modified gravity \citep{Pan_2019}, and the growth history \citep{Alonso_2016}. Measurements of the large-scale velocity field have been performed by ACT \citep{DeBernardis:2016pdv}, SPT \citep{Soergel_2016}, and Planck \citep{Planck:2017xaj}. On the other hand, focusing on inferences on the gas properties, the kSZ effect can be a powerful probe of baryonic processes. One approach to this has been to use the projected field estimator \citep{Hill_2016} that was used to constrain the cosmic baryon fraction. Alternatively, this can be probed by stacking analyses \citep{Schaan_2016,Schaan_2021} which have then been used to understand better the astrophysical processes at play in clusters \citep{Amodeo:2020mmu}. Finally, we note that by combining tSZ and kSZ effects, we can better understand the thermodynamics of a galaxy cluster \citep{Battaglia:2017neq}, including AGN and stellar feedback.

With upcoming high-significance measurements of the tSZ and kSZ effects, we will shift from being limited by statistical uncertainties to theoretical modelling uncertainty. For example, the biggest limitation in cosmological inferences with cluster counts is understanding how to connect the physics of the gas pressure to the mass of the cluster \citep{Hurier:2017jkv}. Likewise, for tSZ and kSZ $N$-point functions, despite the plethora of modelling techniques, including analytic calculations \citep{Shi:2015fua}, semi-analytical models, and fitting the numerical simulations \citep{Efstathiou:2011ac, Trac:2010sp}, cosmological analyses with any of these methods will be limited by our understanding of the distribution of electron pressure and density in the universe. 

In this work, we explore the properties of the electron pressure and density, which encapsulate the astrophysical dependencies of the tSZ and kSZ effects. These cluster properties depend on multi-scale physical processes, such as AGN feedback and plasma physics \citep{Moser:2021llm, DiMatteo:2005ttp, Agertz:2015nzu}, and thus high-resolution hydrodynamic simulations are necessary to study these effects theoretically. We use the IllustrisTNG and SIMBA cosmological simulations, two of the state-of-the-art simulations \citep{Pillepich:2017fcc, Nelson:2017cxy, Marinacci:2017wew, 2018MNRAS.477.1206N, Springel:2017tpz, Nelson:2018uso,Dave_2019}. The huge dynamic range of important processes, from stellar processes to cosmological scales, implies that sub-grid modelling of unresolvable physical processes is necessary, and the freedom in this modelling contributes to the uncertainty in simulation predictions. Whilst our quantitative results are limited to the IllustrisTNG simulation, due to the small volume of the SIMBA simulations, we validate that qualitatively similar features are seen in the SIMBA simulations. This allows us to examine which features are universal. Interestingly we find several important features that are common in both simulations: We find a strong dependence on the density and pressure with concentration, and we find that the radial correlations in these profiles are qualitatively very different between high and low mass halos.

Building on previous investigations \citep{Moser:2021llm, Battaglia:2011cq}, we obtain a pair of parametric formulae for these effects and explore how extensions to previous formulations can improve the accuracy of these formulae. These fitting formulae will have a range of uses: For example, they can be used in the halo-model formalism to compute the tSZ and kSZ $N$-point functions \citep[e.g.][]{Komatsu_2002,Smith_2018} or for the analysis of stacks of clusters for investigations in astrophysical \citep{Amodeo:2020mmu} or cosmological parameters \citep{Silk_1978,Cowie_1978,Kozmanyan_2019}. Alternatively, they can be combined with inexpensive dark-matter only simulations to generate mock observations of the tSZ and kSZ effects  - large numbers of these simulations are necessary for validating experimental analysis techniques and developing new machine-learning-based analysis tools \citep{Thiele:2020zpz,Troster:2019mys}. 

The outline of this paper is as follows: In Section \ref{sec:method}, we present the methodology for finding the fitting formulae for electron pressure and electron density profile using the numerical simulation. In Section \ref{sec:covar}, we discuss the properties of the covariance matrices obtained from the numerical simulation that we used in the fitting. We discuss our analysis of the electron pressure profiles in Section \ref{sec:results_pressure} and the electron density profiles in Section \ref{sec:results_density}. In Section \ref{sec:discuss}, we discuss the implications of our work and give a summary in Section \ref{sec:conclusion}.

\section{Methodology} \label{sec:method}

\subsection{The Sunyaev Zel'dovich Effects}

The tSZ effect is caused by the inverse Compton scattering of CMB photons by hot electrons in the galaxy clusters. The inverse Compton scattering shifts the spectrum of scattered photons to higher frequencies leaving a characteristic spectral feature in the direction of the cluster given by \citep{Zeldovich:1969ff, Sunyaev:1970eu, 1972CoASP...4..173S, 1980ARA&A..18..537S}
\begin{equation}
\frac{\Delta T_{\text{tSZ}}}{T_\text{CMB}} = g\left(\nu\right)y,
\end{equation}
where
\begin{equation}
\begin{split}
g\left(\nu\right) = x \coth{\frac{x}{2}} - 4,
\end{split}
\end{equation}
$x = h \nu/k_b T_{\text{CMB}}$, $h$ is the Planck constant, $\nu$ is the observational frequency, $T_\text{CMB}$ is the CMB temperature, $k_b$ is the Boltzmann constant, 
\begin{equation}
\begin{split}
y = \int \frac{P_e}{m_e c^2} \sigma_T \mathrm{d}s,
\end{split}
\end{equation}
$P_e = n_e k_B T_e$ is the electron pressure, $n_e$ is the electron number density, $T_e$ is the temperature, $m_e$ is the electron mass, $c$ is the speed of light, $\sigma_T = (8\pi/3)(e^2/4\pi \epsilon_0 m_e c^2)^2$ is the Thomson cross-section, $e$ is the electron charge, $\epsilon_0$ is the vacuum permittivity, and $ds$ is the distance along the line-of-sight.

On the other hand, the kSZ effect arises from Compton scattering of CMB photons by electrons with coherent velocities along the line of sight. This results in a Doppler shifting of the CMB photons and generates a temperature anisotropy without altering the photons' spectrum, given by
\begin{equation}
\frac{\Delta T_{\text{kSZ}}}{T_\text{CMB}} = -\frac{\sigma_T}{c} \int n_e \left(\vec{v} \cdot \hat{r}\right) \mathrm{d}s,
\end{equation}
where $\vec{v} \cdot \hat{r}$ is the velocity along the line-of-sight.

The most important physical quantities involved in the tSZ and kSZ effects are the electron pressure and density correspondingly. Thus, to extract cosmological and astrophysical information from these effects, we must have a thorough theoretical understanding of these cluster properties.

\subsection{Hydrodynamical Simulations}

We utilize two simulations for our analysis, including the IllustrisTNG simulations and the SIMBA simulations. We will describe more in details in the following two subsections.

\subsubsection{The IllustrisTNG Simulations}
This analysis is primarily focused on the analysis of the state-of-the-art IllustrisTNG numerical cosmology simulation \citep{Pillepich:2017jle}. We use the large TNG300 simulation, which has a volume of $302.6^3$ Mpc$^3$ and a dark matter mass resolution of $5.9 \times 10^{7}$ $M_\odot$. It tracks 2500$^3$ gas and dark matter particles each. The TNG300 simulation resolves one of the largest amounts of galaxies, and it is one order of magnitude more galaxies compared to its predecessor Illustris simulation. The simulation has the following cosmological parameters: cosmological constant $\Omega_{\Lambda} = 0.6911$, total matter density $\Omega_{m} = 0.3089$, baryon density $\Omega_{b} = 0.0486$, matter fluctuation amplitude $\sigma_8 = 0.8159$, scalar spectral index $n_s = 0.9667$, and Hubble constant $H_0 = 67.74$ $\text{km} \text{s}^{-1}$Mpc$^{-1}$ \citep{Pillepich:2017jle}.

The IllustrisTNG simulation includes a wide range of astrophysical processes, from stellar evolution and gas cooling processes to the impact of supernova winds and AGN feedback. The IllustrisTNG simulation refines the subgrid modelling in the previous Illustris simulation such that it matches better with benchmark observations, such as galaxy stellar mass function and the stellar-to-halo mass relation.

\subsubsection{The SIMBA Simulations}

We make use of the SIMBA simulations to validate our results. The SIMBA flagship run has $100 h^{-1}$ Mpc$^3$ volume and $1024^3$ gas and dark matter elements each, with a dark matter particle mass resolution of $9.6 \times 10^{7}$ $M_\odot$. The simulation has the following cosmological parameters: $\Omega_{\Lambda} = 0.67$, $\Omega_{m} = 0.3$, $\Omega_{b} = 0.048$, $\sigma_8 = 0.82$, $n_s = 0.97$, and $H_0 = 68$ kms$^{-1}$Mpc$^{-1}$.

The SIMBA simulation also includes multi-scale physical processes such as AGN jets, radiative winds, X-ray feedback, and $H_2$-based star formulation. The simulation matches with a wide range of observations such as AGN luminosity functions, galaxy color, and stellar-to-halo mass bimodality \citep{Dave_2019}. The different subgrid modelling makes the comparison with the IllustrisTNG simulations a useful probe of the generality of our results beyond a single simulation.

A crucial part of our analysis is the accurate computation of the cluster profile covariance matrix (Section \ref{sec:covar}), and this can only be performed when we have more objects than the number of radial bins. The smaller box size of the SIMBA simulation means there are fewer halos ($\sim 1/8^{\text{th}}$ the number of objects), especially at high masses. Therefore, we have a significantly reduced range of mass and redshifts bins that we can analyze quantitatively with the SIMBA simulations. As a result, it is difficult to accurately constrain the mass and redshift evolution of the electron pressure and density. In addition, we do not have a gravity-only counterpart of the SIMBA simulation box available. This means that halo identification is different (for the IllustrisTNG simulation, the halos were found in the gravity-only run in order to facilitate straightforward evaluation of analytic models that typically rely on gravity-only calibrated halo mass functions). For this reason, the halo masses are expected to differ between SIMBA and IllustrisTNG, although not at a level important for the conclusions drawn in this work. Despite these limitations, we found that the SIMBA simulations are very useful for cross-checking the results of the analysis of the IllustrisTNG simulations qualitatively.

\subsection{Data Extraction and Modelling}

Using the Rockstar catalog of (gravity-only) IllustrisTNG parent halos \citep{Behroozi:2011ju} produced in \citet{Gabrielpillai_2021}, we measure properties of the individual halos, such as the radius $R_{200}$ and mass $M_{200}$, where $R_{200}$ is the radius at which the mean interior density is 200 times the critical density $\rho_{cr}(z)$ \citep{Battaglia:2011cq}, defined as
\begin{equation}
\rho_{cr}\left(z\right) = \frac{3H_0^2}{8\pi G}\left[\Omega_m \left(1+z\right)^3 + \Omega_{\Lambda}\right],
\end{equation}
where $z$ is the redshift and $G$ is the gravitational constant.

Using the hydrodynamic data fields available in the IllustrisTNG simulation, we compute the required volume-weighted $P_e$ and $n_e$ for each gas particle $i$ as
\begin{align}
V_i P_{e,i} &= x_i m_i \epsilon_i \frac{4 X_H (\gamma-1)}{1 + 3 X_H + 4 X_H x_i}\,,\\
V_i n_{e,i} &= x_i m_i \frac{X_H}{m_p}\,,
\end{align}
where $x$ is the electron abundance, $m$ the mass, $\epsilon$ the internal energy,
$X_H=0.76$ the primordial hydrogen mass fraction, $\gamma=5/3$ the adiabatic index, and $m_p$ the proton mass. The particles are then grouped into radial bins $b_\alpha$ in which the averages are computed:
\begin{equation}
A_\alpha = V^{-1}(b_\alpha) \sum_{i \in b_\alpha} V_i A_i\,,
\end{equation}
where $A = P_e, n_e$. The radial bins are chosen as 27 logarithmically spaced intervals between 0.04 and 1.34 $R_{200}$.

We further separate the halos into different mass, redshift, and concentration bins. For mass $M$, we group data into bins of width $10^{0.2}$ $h^{-1} M_\odot$, starting with the first bin $10^{13}$ - $10^{13.2}$ $h^{-1} M_\odot$. We strike a balance between computational time and including a representative amount of data. Consequently, the redshift bins are chosen as the 20 full IllustrisTNG snapshots \citep{Pillepich:2017jle}. After that, we further separate the halos evenly into four different concentration groups, namely the lowest (0th - 25th percentile), low (25th - 50th percentile), high (50th - 75th percentile), and highest (75th - 100th percentile). In each mass, redshift, and concentration bin, we normalize the electron pressure (electron density) by the self-similar amplitude for pressure $P_{200}$ (density $n_{200}$), defined as
\begin{equation}
\begin{split}
P_{200} = \frac{200 GM_{200}\rho_{cr}}{2R_{200}} \frac{\Omega_b}{\Omega_m}, \\
n_{200} = \frac{200\rho_{cr}}{X_H m_p} \frac{\Omega_b}{\Omega_m}.
\end{split}
\end{equation}
Then, we calculate the average profile $\bar{P}_e = \left \langle P_e/P_{200} \right \rangle$ ($\bar{n}_e = \left \langle n_e/n_{200} \right \rangle$).

We look for the best-fitted generalized Navarro–Frenk–White (NFW) profile \citep{Navarro:1996gj}
\begin{equation}
\begin{split}
\bar{P}_e\left(x\right)  = P_0 \left( \frac{x}{x_c} \right)^\gamma \left[1+\left(\frac{x}{x_c}\right)^\alpha \right]^{-\beta}, \\
\bar{n}_e\left(x\right)  = n_0 \left(\frac{x}{x'_c}\right)^\gamma \left[1+\left(\frac{x}{x'_c}\right)^\alpha\right]^{-\beta'},
\end{split}
\label{eqn:10}
\end{equation}
where $P_0$, $n_0$, $x_c$, $x'_c$, $\beta$, and $\beta'$ are the parameters for the fitting. As suggested in \citet{Nagai:2007mt} and \citet{Arnaud:2009tt}, we use the fixed value $\alpha = 1.0$ and $\gamma = -0.3$, while $\alpha$ and $\gamma$ are also allowed to vary in general. Our base analysis, consistent with past work \citep{Battaglia:2011cq}, considers the evolution of these parameters as
\begin{align}
A = A_0 \left(\frac{M}{10^{14} M_\odot}\right)^{\alpha_m} \left(1+z\right)^{\alpha_z},
\end{align}
where $A \in  \{P_0, n_0, x_c, x'_c, \beta, \beta' \}$ and $A_0$, $\alpha_{m}$, and $\alpha_{z}$ are the parameters that are fit for.

In this work, we also consider an extended parameterization, where we investigate the importance of deviations from a power law in mass and the inclusion of concentration dependence. Thus $P_0(M, z, c)$, $n_0(M, z, c)$, $x_c(M, z, c)$, $x'_c(M, z, c)$, $\beta(M, z, c)$, and $\beta'(M, z, c)$ are additional functions of concentration $c$, and have the generic form \citep{Battaglia:2011cq}
\begin{equation}
A = A_0 (1+z)^{\alpha_z}
\left(\frac{c}{10.0}\right)^{\alpha_c}
\times
\begin{cases}
\left(\frac{M}{M_\text{cut}}\right)^{\alpha_m}, M < M_\text{cut}\\
\left(\frac{M}{M_\text{cut}}\right)^{\alpha'_m}, M > M_\text{cut}
\end{cases},
\label{eqn:11}
\end{equation}
where $A$ is as above and we also fit for $M_{\text{cut}}$,  $\alpha'_{m}$, and $\alpha_{c}$. Our definition of concentration is $c = R_\mathrm{200}/R_\mathrm{scale}$, where $R_\mathrm{scale}$ is the Klypin scale radius \citep{Klypin_2011}.  Note that the parameters for the electron pressure ($P_0, x_c, \beta$) share the same $M_{\text{cut}}$ and likewise for the electron density ($n_0, x'_c, \beta'$).

These proposed extensions to the \citet{Battaglia:2011cq} parameterization are physically well-motivated: $c$ traces the mass accretion histories, and thus it is a natural parameter to characterize the halo. This is further motivated by \citet{Wadekar_2022}, who found evidence that the integrated $y$ signal exhibits a dependence on $c$ in the IllustrisTNG simulations. Also, we notice that there is only a weak dependence of $c$ with mass. For example, in the $M = 10^{13.0}-10^{13.2} h^{-1}$ $M_\odot$ and $z=0.0$ bin, we have a mean concentration $\overline{c} = 5.65$ with a sample standard deviation 1.99, while in the $M = 10^{14.0}-10^{14.2} h^{-1}$ $M_\odot$ and $z=0.0$ bin, we have $\overline{c} = 4.53$ and a sample standard deviation 1.47, in accordance with the results in \citet{Child:2018skq}, suggesting that it should be treated as a separate parameter. As discussed later, it is found to be important in our results. On the other hand, the broken power law (BPL) in halo mass is motivated by the fact that AGN and supernova feedback impact lower and higher mass halos in qualitatively different ways and have been seen in the integrated Compton-$y$ signal in previous simulations \citep{LeBrun2017} and analytical models \citep{Shaw_2010}. Further, recent measurements by ACT and the Dark Energy Survey suggest there are observed deviations from a power law in the observed profiles \citep{Pandey_2021}.

We describe the likelihood function by a log-normal distribution
\begin{equation}
\log{\mathcal{L}} = \sum_{\text{bins}} -\frac{1}{2}\left(\log \vec{d}- \log \vec{\mu}\left(\vec{\theta}\right)\right)^\mathsf{T} \Sigma^{-1} \left(\log \vec{d}-\log \vec{\mu}\left(\vec{\theta}\right)\right),
\end{equation}
where $\vec{d}$ is the electron pressure or electron density data vector extracted from the IllustrisTNG numerical simulation as described above, $\vec{\mu}(\vec{\theta})$ is the generalized NFW model (Eq. (\ref{eqn:10})) with parameters $\vec{\theta}$, and $\Sigma$ is the covariance matrix. A log-normal distribution is appropriate because the electron pressure and density cannot be negative, and studies on the total integrated $y$ signal of a cluster have shown it follows a log-normal distribution \citep{Stanek_2010,Kay_2012}. We also apply the Hartlap factor for the inverse covariance matrix (precision matrix) \citep{Hartlap:2006kj}. The summation goes over each mass, redshift, and concentration bin. With the 27 logarithmically spaced radial bins we have, we include mass, redshift, and concentration bins that have more than 29 halos to ensure an invertible covariance matrix, which sets an upper bound for high mass or high redshift. The amount of radial bins is chosen to save computational time and to include more high mass (up to $M = 10^{14}$ $h^{-1} M_\odot$) and high redshift halos (up to $z=2.0$) into the fitting, which have an intrinsically lower population in the simulation. 

We use the Markov chain Monte Carlo (MCMC) method with the implementation given in the \emph{emcee} Python package \citep{2013PASP..125..306F} to find the best-fit parameters. A flat prior that covers a large parameter space is used.

\section{The evolution of radial correlations} \label{sec:covar}

\begin{figure}
\centering
\plottwo{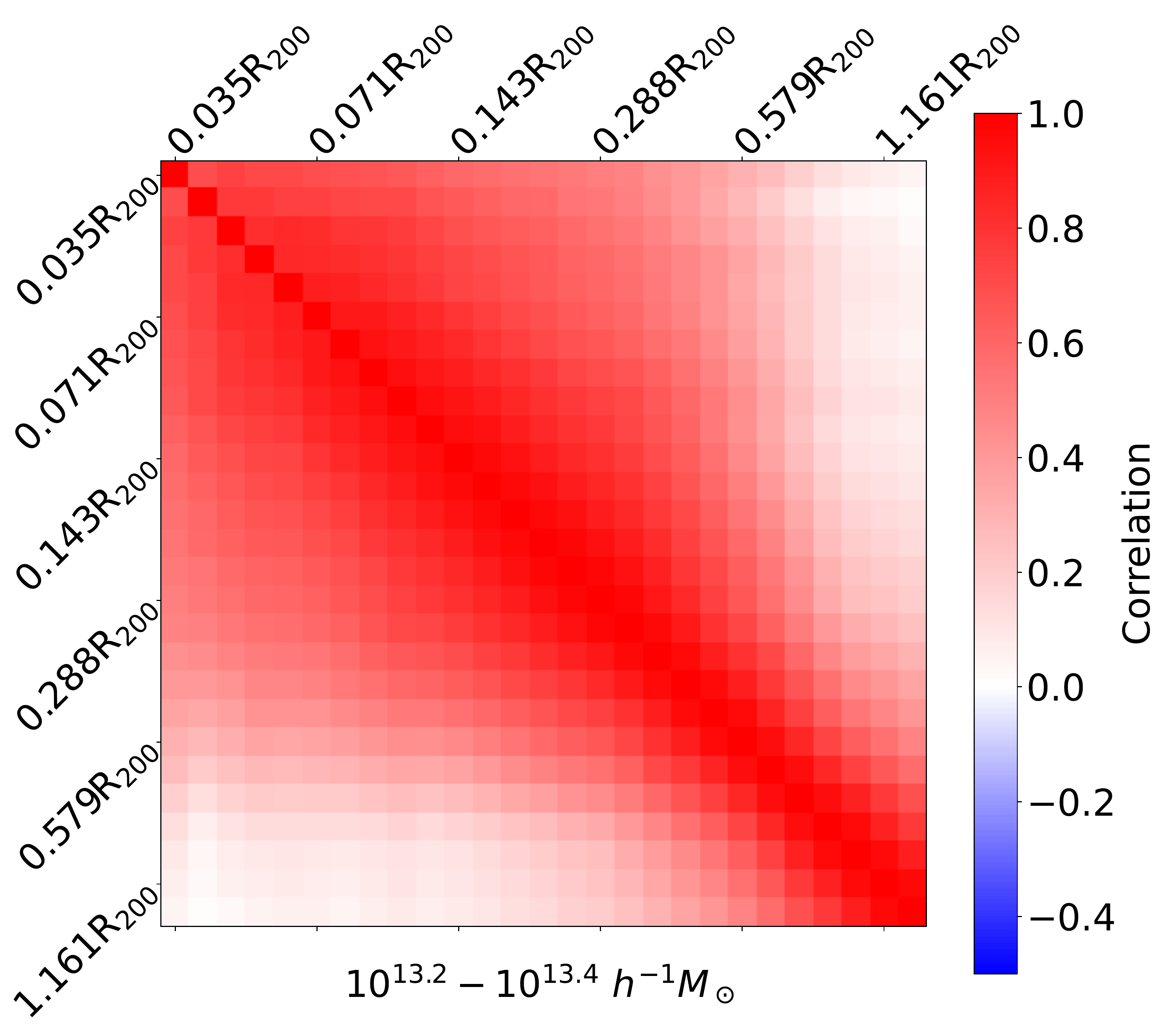}{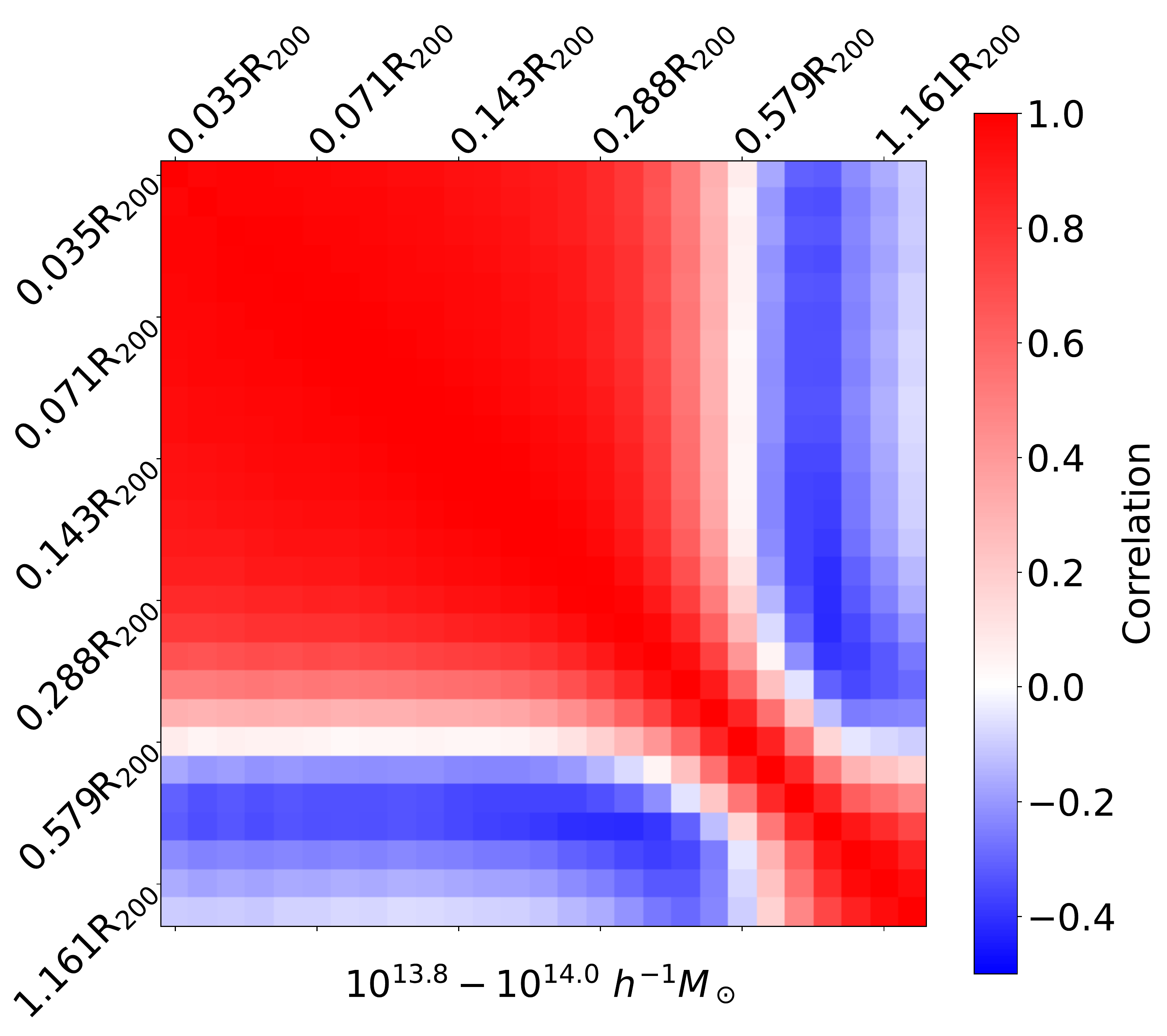}
\plottwo{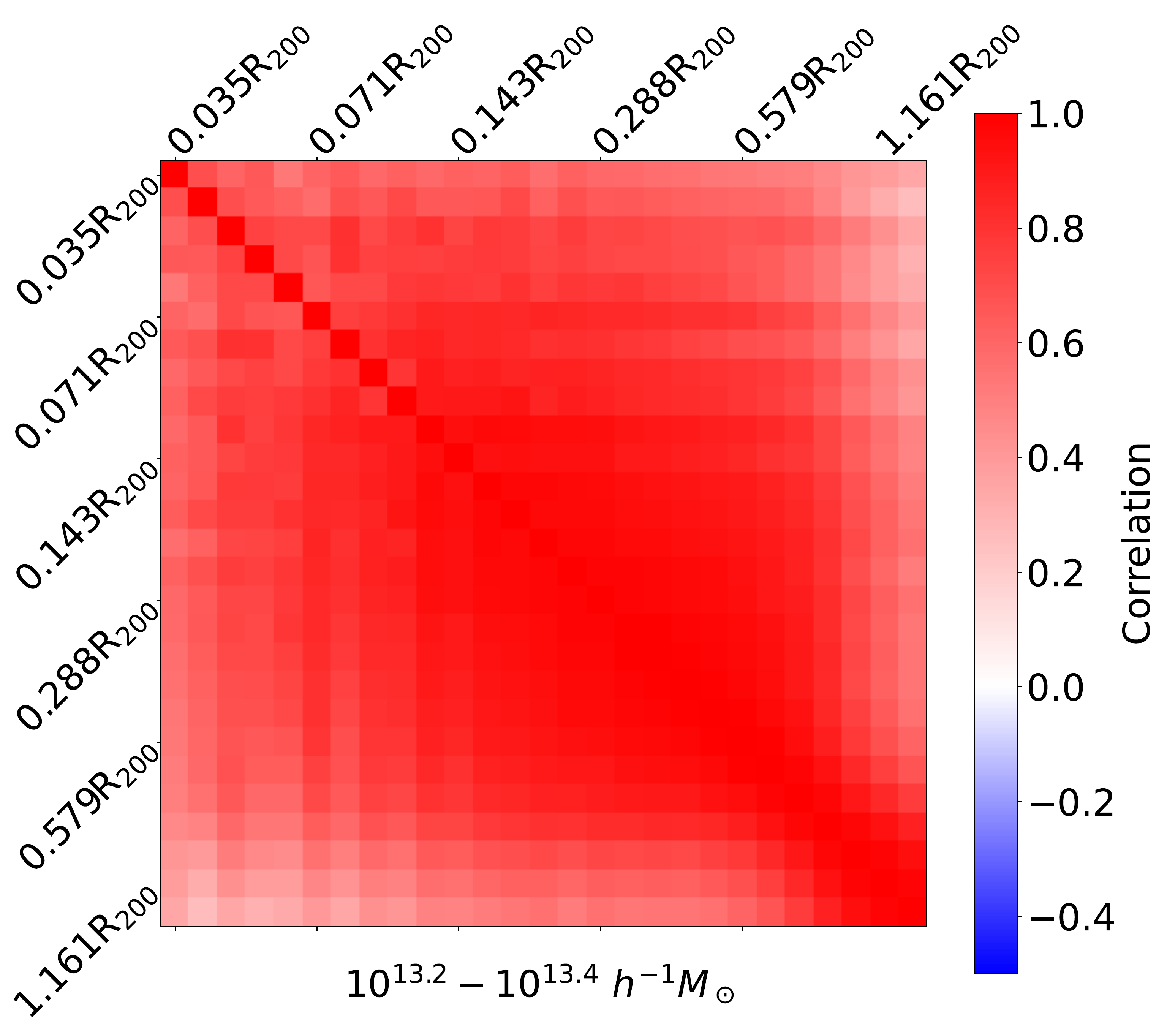}{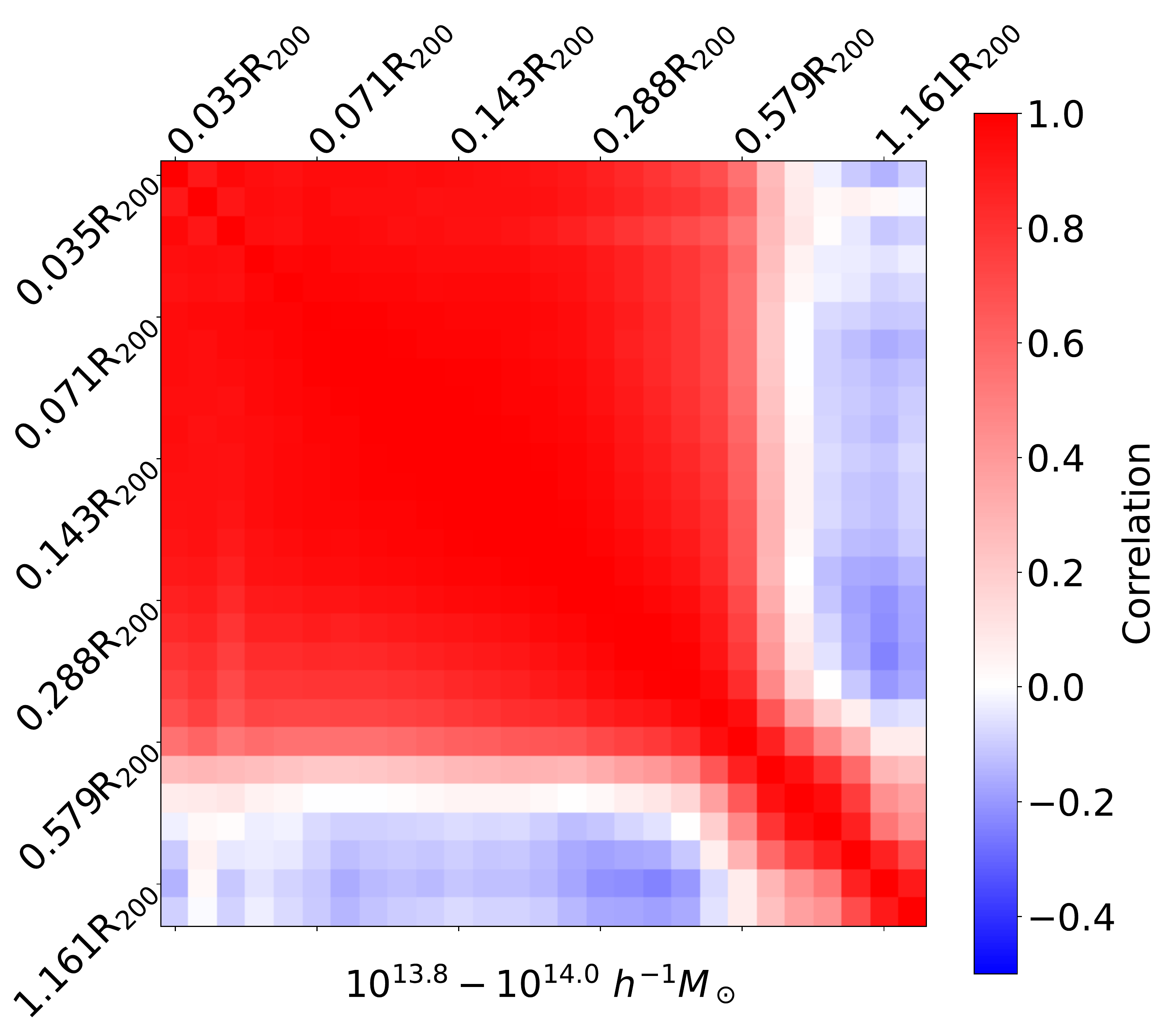}
\caption{The correlation matrix for electron pressure in the $10^{13.2}-10^{13.4}$ $h^{-1} M_\odot$ (left) and $10^{13.8}-10^{14.0}$ $h^{-1} M_\odot$ (right) mass bins with $z = 0.0$ for the IllutrisTNG (top) and SIMBA (bottom) simulations. Contrary to the expectation of self-similar models, the correlation matrix across different mass bins have a different structure.}
\label{fig:0}
\end{figure}
An important ingredient in our analysis is the covariance matrix $\Sigma$. The covariance matrix shows how different radial bins are correlated and so determines how many effective bins we have. Self-similar models of cluster properties have been successful at capturing the broad properties of simulations and observations \citep{Vikhlinin_2006,Kravtsov_2012,Mantz_2016,McDonald_2017}. However, in our investigation of the structure of the electron pressure and density covariance matrices, we find strong deviations from self-similarity that we attribute to the different impact of feedback processes in the cluster.

In the top panel of Fig. \ref{fig:0}, we show the correlation matrix in the low and high mass bins at redshift 0.0 in the IllustrisTNG simulation. In the low mass halos, all radial bins are slightly positively correlated. On the other hand, the inner region ($r < 0.3 R_{200}$) of the high mass halo is highly positively correlated, and the middle region of the halo ($r \sim 0.6 R_{200}$) is negatively correlated with the inner region. The strong correlations in the high-mass halos are thought to arise due to the very strong feedback in massive halos at low redshift. The feedback in the center heats up and expels matter from the halo center leading to strong correlations in the center. In the bottom panel of Fig. \ref{fig:0}, we plot the corresponding plots for the SIMBA simulations and find qualitatively similar results indicating that this effect is not a feature of the exact subgrid models used in IllustrisTNG. This suggests the reported effect here is likely important to create simulation outputs that are consistent with observations.

The high correlation over distant radial bins also motivates us to use a comparatively smaller number of bins (27 bins) as the data vector, which helps to include larger mass and redshift bins that have a smaller number of halos. A potential cause for concern could be the small number of objects used to compute some of the covariance matrices. We can assess their stability by examining how the covariance matrices change between adjacent bins, which typically have vastly different numbers of objects. We generally observe rather gradual transitions, supporting our view that the covariance matrices are stable enough.

\section{Electron pressure} \label{sec:results_pressure}

\begin{figure}
\centering
\includegraphics[width=1.0\textwidth]{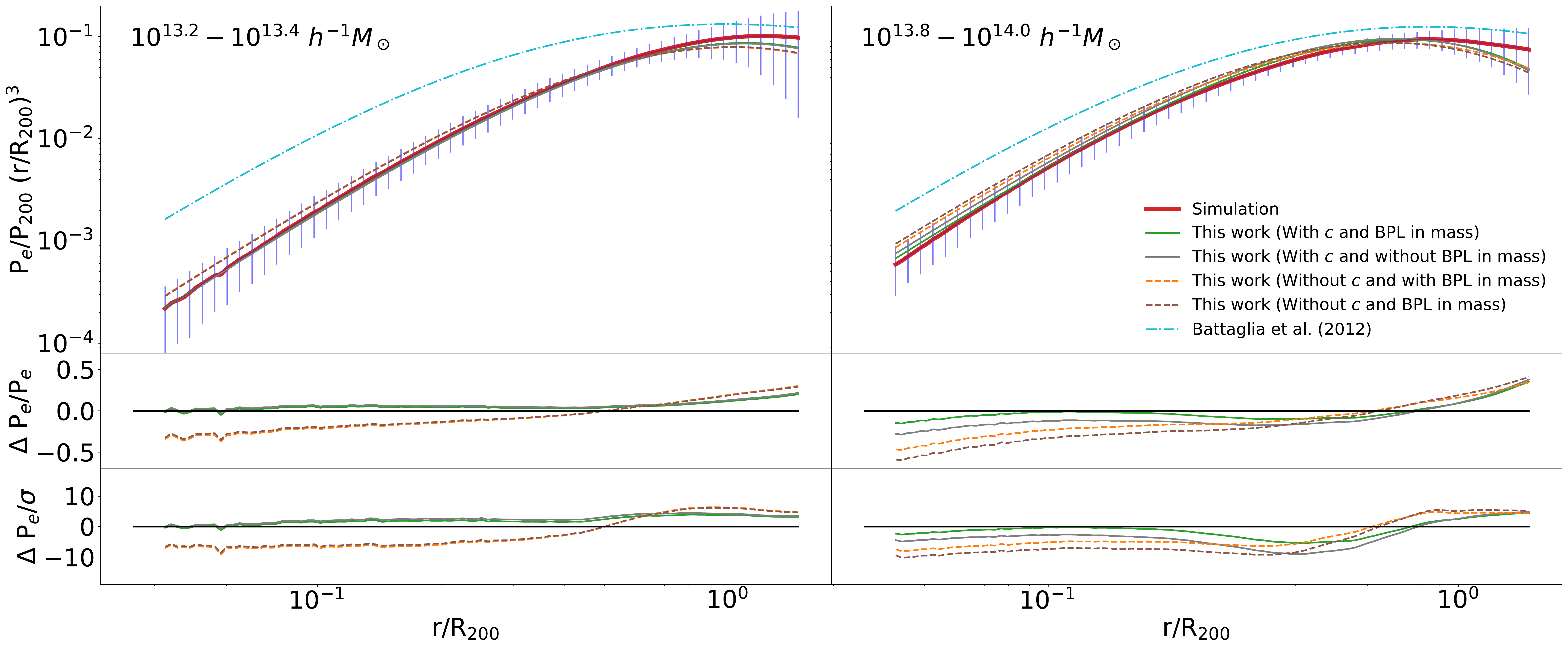}
\caption{Example fits for the electron pressure in the $10^{13.2} - 10^{13.4}$ $h^{-1} M_\odot$ (left) and $10^{13.8} - 10^{14.0}$ $h^{-1} M_\odot$ (right) bin with $z = 0.0$ and lowest concentration in both cases. The different color lines indicate different cases where the $c$ dependence or broken power law (BPL) in mass is included or not. The blue bars show the standard deviation in different radial bins (note these are not the errors on the mean pressure, which are smaller by the square root of the number of halos in the bin $\sqrt{N}$). The cyan line uses the parameters in \citet{Battaglia:2011cq}. The lower panels show the deviation compared to the mean (middle) and the error on the mean (bottom).
}
\label{fig:2}
\end{figure}

In the left and right panels of Fig. \ref{fig:2}, we show the electron pressure profiles for the lower mass $10^{13.2} - 10^{13.4}$ $h^{-1} M_\odot$ and higher mass $10^{13.8}-10^{14.0}$ $h^{-1} M_\odot$ bin with  $z=0.0$ and the lowest concentration bin in both cases. The red line shows the average profile in the bin, and the blue bars denote the sample standard deviation. We first compare our measurements to the fits obtained in \citet{Battaglia:2011cq}. We find that in the high mass bin, the pressure profiles in the IllustrisTNG simulation show reasonable agreement with the results from \citet{Battaglia:2011cq}. The differences seen arise from the different implementations of the baryonic processes, particularly the AGN and supernova feedback, and thus are useful to understand the theoretical uncertainty of these properties. At lower masses, and especially lower radii, we see highly significant differences. The mass resolution of the simulations used in \citet{Battaglia:2011cq} is a factor of $\sim 100$ lower than that of our simulations, and thus large differences are expected as we approach the mass resolution at which the  \citet{Battaglia:2011cq} formula was calibrated.

Using the same formalism in \citet{Battaglia:2011cq}, we refit the electron pressure profile. The results of this are shown by the brown line in Fig. \ref{fig:2}. Even from direct observation, it is clear that the fitted curve deviates from the simulation data significantly. To put it quantitatively, we show the percentage difference between the best-fit model with the average electron pressure profile obtained from the simulation in the middle panel of Fig. \ref{fig:2}, and in the bottom panel, we show the ratio between the difference between the best-fit model and the data $\Delta P_e$ and the standard error of the average electron pressure profile $\sigma$. It gives a rough estimation of the goodness of fit. The electron pressure can deviate by 60\% ($\sim 10 \sigma$) in the inner part of the halo for the higher mass case.

It can also be seen that the outer boundary ($r > R_{200}$) exhibits a larger standard deviation, which is expected as it is more significantly affected by in-falling material and nearby substructures and clusters. Nonetheless, they will not contribute significantly to the fitting as they will be down-weighted by the covariance matrix. In our analysis, we examined three different choices of radial binning cutoff, including 1.01 $R_{200}$, 1.34 $R_{200}$, and 1.76 $R_{200}$. Whilst extending to larger radii leads to a very large deviation in the fitted parameters, reducing the range has less impact. This indicates that our chosen range is sufficient. Our choice is consistent with the result of \citet{Moser:2021llm}, who investigates the relative importance of the one and two halo terms systematically. We further verified that changes to the maximum radius do not resolve the poor fit seen here (this robustness remains when we consider extended parameterizations). 

Given these large residuals, we consider how the fit changes when we include our first extension, the BPL in mass in $P_0$; this is shown as the orange dotted line in Fig. \ref{fig:2}. Whilst there is a general improvement to the profile, especially in the higher mass halos, we still see large differences ($\sim 47\%$ in the inner region) in the profiles. Next, we consider extending the original model by including a $c$ dependence. The results of our best fit model are shown as gray lines in Fig. \ref{fig:2}. Generally, we find that this provides a better fit than the BPL model. However, there are still large residuals, particularly at higher mass -- further indicating that the single power law in mass is not sufficient. This leads us to consider our full model, including both $c$ dependence and the BPL in the $P_0$ parameter, and the fit with this model is represented by the green line in Fig. \ref{fig:2}. With $c$ dependence and BPL in mass for $P_0$, the absolute percentage difference throughout the halo is well below 20\% for the lower mass case, or equivalently the absolute ratio stays lower than 3.8$\sigma$, compared to 8.6$\sigma$ without these features.

\begin{figure}
\centering
\includegraphics[width=1.0\textwidth]{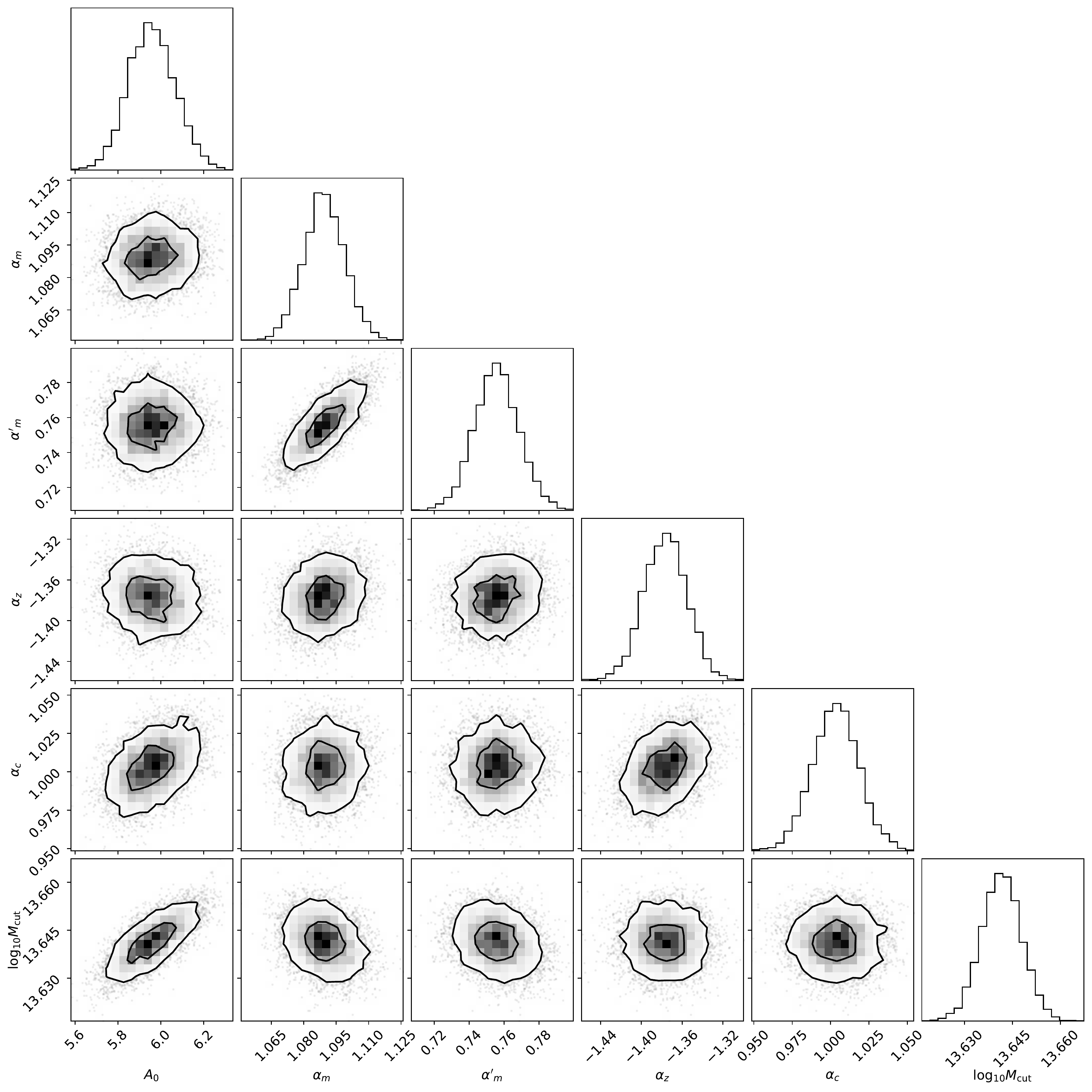}
\caption{Posterior probability distribution for the parameters of the electron pressure profiles, Eq. (\ref{eqn:10}) and (\ref{eqn:11}), with 1$\sigma$ and 2$\sigma$ contours. We plot the parameters describing the change in the amplitude of the pressure profile $P_0$ with mass, redshift, and concentration, including the reference amplitude $A_0$, its evolution with redshift $\alpha_z$, concentration $\alpha_c$, and mass below $\alpha_m$ and above $\alpha_m'$ the mass break $M_{\text{cut}}$, where the difference in values shows the importance of the mass break. $\alpha_m, \alpha_m', \alpha_z,$ and $\alpha_c$ all depart from zero showing the importance to include mass, redshift, and concentration as parameters.}

\label{fig:1}
\end{figure}

The corner plot in Fig. \ref{fig:1} shows the posterior probability distribution of the fitting parameters for the pressure profiles, Eq. (\ref{eqn:10}) and (\ref{eqn:11}), for our model with $c$ dependence and the BPL (green line in Fig. \ref{fig:2}). We plot the parameters describing the change in the amplitude of the pressure profile $P_0$ with mass, redshift, and concentration as an illustration, including the reference amplitude $A_0$, its evolution with redshift $\alpha_z$, concentration $\alpha_c$, and mass below $\alpha_m$ and above $\alpha_m'$ the mass break $M_{\text{cut}}$, where the difference in values shows the importance of the mass break. The best-fit values are listed in Table \ref{table:1}, where the upper (lower) bound uncertainty is given by the absolute difference between the 84th (16th) and 50th percentiles. We find a significant dependence of the pressure profile with mass ($\alpha_m$ and $\alpha_m'$) and redshift ($\alpha_z$), in accord with the results given in \citet{Battaglia:2011cq}. For example, $P_0$ has $\alpha_z = -1.38^{+0.02}_{-0.02}$, which shows that $\bar{P_{e}}$ is smaller when the redshift is higher. In addition, $c$ also contributes significantly to the electron pressure profile. For example, $P_0$ has $\alpha_c = 1.00^{+0.01}_{-0.02}$, showing that a more concentrated halo has a higher $\bar{P_{e}}$. $\alpha_m$, $\alpha_m'$, $\alpha_z$, and $\alpha_c$ all depart from zero (i.e., no dependence on that parameter) by a high significance indicating the dependence on these variables is important, and ignoring them would lead to a bad fit, which will be further discussed later quantitatively. While some of the parameters show a strong correlation, such as $A_0$ of $P_0$ with $M_{\text{cut}}$, which is an intrinsic property of Eq. (\ref{eqn:11}), the weak correlation between $\alpha_c$ of $P_0$ and $A_0$, $\alpha_m$, $\alpha'_m$, and $\alpha_z$ of $P_0$ is an interesting observation because we expect that the stacked $c$ follows a function of mass and redshift \citep{2012MNRAS.423.3018P, Bullock:1999he, Child:2018skq}. It suggests that the intrinsic scatter of $c$ in each mass and redshift bin is also significant, and $c$ should be standalone and form one extra parameter to the fitting.

\begin{table}
\centering
\begin{tabular}{c|c|c|c|c|c|c}
 & $A_0$ & $\alpha_m$ & $\alpha_m'$ & $\alpha_z$ & $\alpha_c$ & \multicolumn{1}{c}{$\log_{10}M_{\text{cut}}$ ($h^{-1}M_\odot$)} \\ \hline
$P_0$     & $6.0_{-0.1}^{+0.1}$     & $1.09_{-0.01}^{+0.01}$          & $0.76_{-0.01}^{+0.01}$           & $-1.38_{-0.02}^{+0.02}$          & $1.00_{-0.02}^{+0.01}$          & \multirow{3}{*}{$13.64_{-0.01}^{+0.01}$}                    \\ 
$x_c$     & $1.02_{-0.02}^{+0.02}$     & $-0.29_{-0.02}^{+0.02}$          & --           & $-0.33_{-0.04}^{+0.04}$          & $-1.24_{-0.03}^{+0.03}$         &                                       \\ 
$\beta$   & $6.3_{-0.1}^{+0.1}$     & $0.01_{-0.01}^{+0.01}$          & --           & $-0.07_{-0.03}^{+0.03}$         & $-0.82_{-0.03}^{+0.03}$         &                                       \\ 
\end{tabular}
\caption{Best-fit parameters for the electron pressure profiles, Eq. (\ref{eqn:10}) and (\ref{eqn:11}), when concentration dependence and a broken power law (BPL) in mass in $P_0$ are included.}
\label{table:1}
\end{table}

\begin{figure}
\centering
\plottwo{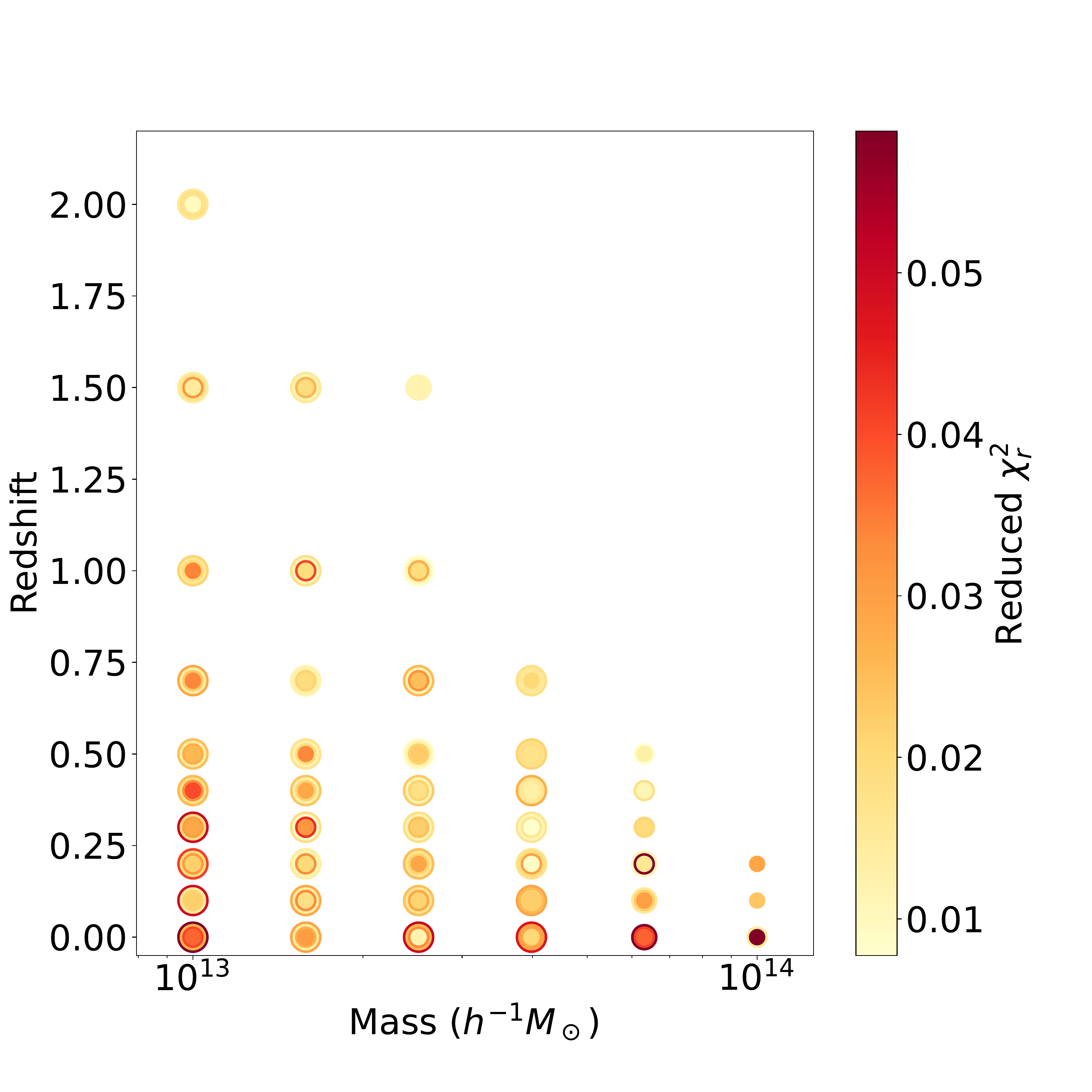}{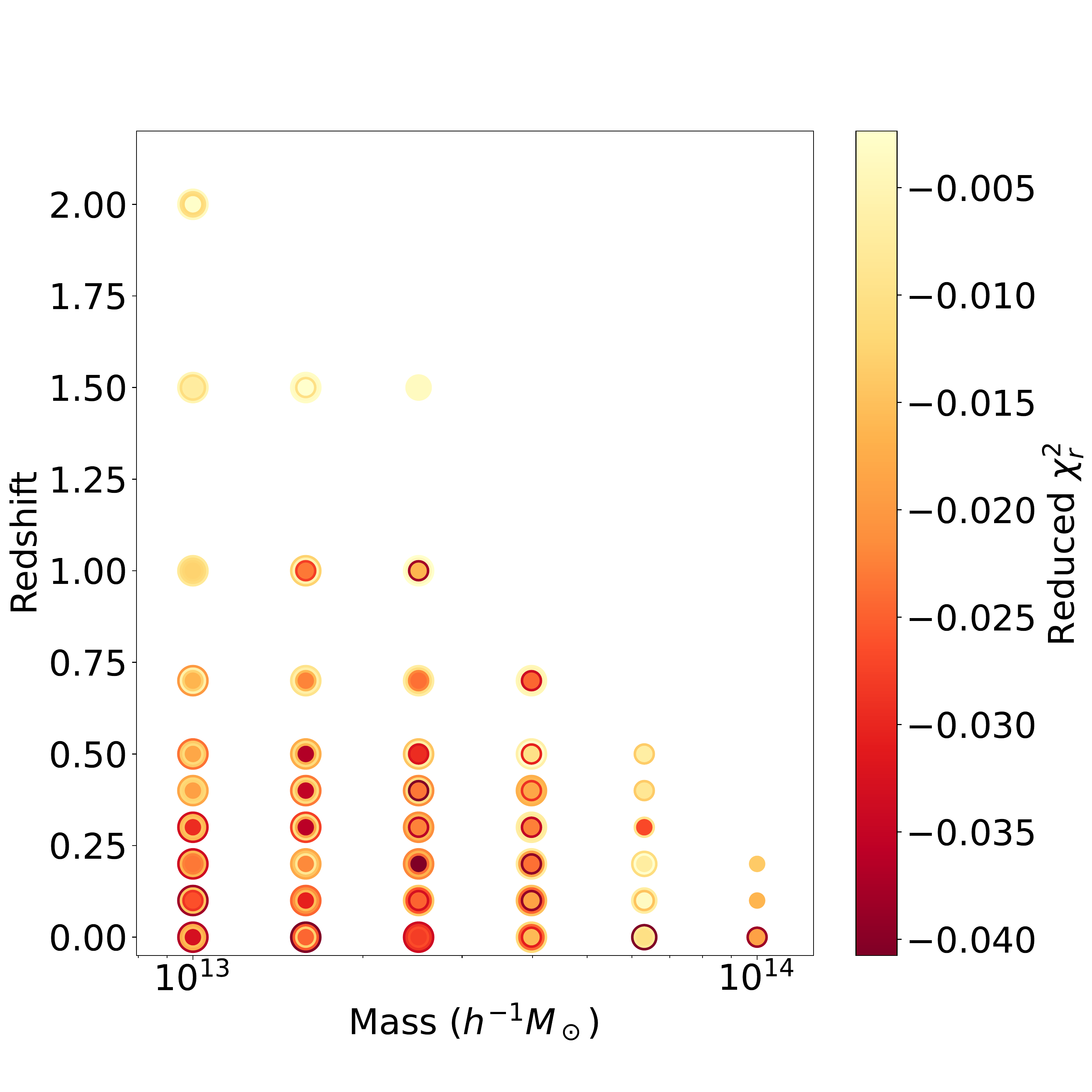}
\caption{(Left) The reduced chi-square statistics and (right) the decrease in reduced chi-square statistics in different mass, redshift, and concentration bins when halo concentration dependence and broken power law (BPL) in mass are included. In each mass and redshift, the outermost circle indicates the highest concentration, and the innermost circle indicates the lowest concentration bin. The degree of freedom is $\sim 4200$.}
\label{fig:3}
\end{figure}

To further examine the robustness of the fitting, we plot the reduced chi-square statistics of different mass, redshift, and concentration bins in Fig. \ref{fig:3}. For each mass and redshift, there is a concentric circle with four different colors that represents the four different concentration bins, where the outermost circle is for the highest concentration, and the innermost circle is for the lowest concentration. In the high redshift region, the reduced $\chi^2_r$ is generally smaller than the low mass or low redshift region. For example, in the $10^{13.2}$ - $10^{13.4}$ $h^{-1} M_\odot$, redshift 0.0, and lowest concentration bin, which is the same as the left panel of Fig. \ref{fig:1}, we have a reduced chi-square $\chi^2_r = 0.0308$, and in the $10^{13.2}$ - $10^{13.4}$ $h^{-1} M_\odot$, redshift 1.5, and lowest concentration bin, we have a reduced chi-square $\chi^2_r = 0.0195$.

There are several possible reasons for this bad fitting in the low mass, low redshift region. We first examine whether our log-normal likelihood is accurate or not. We perform a by eye comparison of the observed distribution within a single radial bin when compared to a log-normal distribution fit to that bin alone. We generally find good agreement between the log-normal distribution and our data. However, some bins show evidence of large outliers. These outliers likely represent the impact of mergers or violent events in the clusters. This is supported by the work of \citet{Yu_2015}, who found that recent mergers lead to large scatter in the Compton-$y$, especially at large radii. Manually removing all these events is prohibitive given the large number of clusters we consider. However, we verified that removing these objects by using a deviation clipping led to minimal changes in our fits. Thus indicating that these are likely not the cause of the poor fits. There are other possible reasons, including that we may be missing some important parameters besides mass, redshift, and concentration to describe the electron pressure profile. Note that the BPL in mass indicates a cutting mass of $10^{13.6} \sim 4 \times 10^{13}$ $h^{-1} M_\odot$. This cutting mass could indicate that additional physical processes, such as AGN feedback, are important when $M < 10^{13.6}$ $h^{-1} M_\odot$. Indeed, studies are showing that when feedback effects are included, the generalized NFW profile does not produce a good fit \citep{2021MNRAS.504.5131L}, which could explain the bad fit we have. It also suggests that fitting the generalized NFW to data may not be the right choice for high signal-to-noise ratio measurements. 

The reduced chi-square statistic allows us to examine the improvement of our extended model, compared to the base model, across mass and redshift. In the right panel of Fig. \ref{fig:3}, we plot the changes of reduced chi-square when we include the mass break and $c$ dependence. We obtained a significant reduction in chi-square in all different bins. For example, in the $10^{13.2}$ - $10^{13.4}$ $h^{-1} M_\odot$, redshift 0.0, and lowest concentration bin, we have a 0.0308 decrease in the reduced chi-square. In the high mass, high redshift bins, such as the $10^{13.6}$ - $10^{13.8}$ $h^{-1} M_\odot$, redshift 0.7, and lowest concentration bin, we have a 0.0245 decrease in the reduced chi-square. Overall, we have a chi-square $\chi^2_r = 3.29 \gg 1$, indicating the one-halo generalized NFW profile is a bad fit. Nonetheless, the large decrease implies that even though we still do not have a good fit, the introduction of the $c$ parameter can significantly improve the fitting.

The reduced chi-square statistic also allows us to evaluate the importance of our two new features: the BPL in mass and the $c$ parameter. In Table \ref{table:2}, we list the decrease in the reduced chi-square when we include individual features in the fitting. First, we see that having a mass break and without $c$ dependence reduces the reduced chi-square by a max of $0.23$. Including the $c$ dependence, without the BPL in mass, leads to a greater reduction of $1.74$. Given the number of degrees of freedom of $\sim 4200$, both of these reductions warrant the inclusion of these new features. Finally, we consider the value of having both $c$ dependence and a BPL in mass; this leads to a greater reduction than either feature alone, $2.06$! However, we see that the reduction in chi-square is similar if we have a BPL in just $P_0$ or all the parameters. Thus our favoured model, and the one used in Fig. \ref{fig:2}, has a BPL only in $P_0$ and the $c$ dependence.

\begin{table}
\centering
\begin{tabular}{lllllllll}
                                            &                           & \multicolumn{7}{c}{Broken power law in}                                                                                                                                                                                                     \\
\multicolumn{1}{l|}{}                       & \multicolumn{1}{l|}{None} & \multicolumn{1}{l|}{$x_c$} & \multicolumn{1}{l|}{$P_0$} & \multicolumn{1}{l|}{$\beta$} & \multicolumn{1}{l|}{$x_c$ and $P_0$} & \multicolumn{1}{l|}{$x_c$ and $\beta$} & \multicolumn{1}{l|}{$P_0$ and $\beta$} & $x_c$, $P_0$, and $\beta$ \\ \hline
\multicolumn{1}{l|}{With concentration}     & \multicolumn{1}{l|}{-1.74}    & \multicolumn{1}{l|}{-1.96}     & \multicolumn{1}{l|}{-2.01}     & \multicolumn{1}{l|}{-1.93}       & \multicolumn{1}{l|}{-2.01}               & \multicolumn{1}{l|}{-2.05}                 & \multicolumn{1}{l|}{-2.00}                 & -2.06                         \\
\multicolumn{1}{l|}{(and free $\gamma$ {[}Section 5.2{]})}     & \multicolumn{1}{l|}{(-2.65)} & \multicolumn{1}{l|}{(-2.87)}    & \multicolumn{1}{l|}{(-2.90)}     & \multicolumn{1}{l|}{(-2.84)}     & \multicolumn{1}{l|}{(-2.90)}       & \multicolumn{1}{l|}{(-2.98)}               & \multicolumn{1}{l|}{(-2.91)}                 & (-2.99)    \\ \hline
\multicolumn{1}{l|}{Without concenteration} & \multicolumn{1}{l|}{--}    & \multicolumn{1}{l|}{-0.11}     & \multicolumn{1}{l|}{-0.17}     & \multicolumn{1}{l|}{-0.08}       & \multicolumn{1}{l|}{-0.20}               & \multicolumn{1}{l|}{-0.23}                 & \multicolumn{1}{l|}{-0.22}                 & -0.23     \\
\multicolumn{1}{l|}{(and with free $\gamma$ {[}Section 5.2{]})} & \multicolumn{1}{l|}{(-0.80)}    & \multicolumn{1}{l|}{(-0.92)}     & \multicolumn{1}{l|}{(-0.96)}     & \multicolumn{1}{l|}{(-0.89)}       & \multicolumn{1}{l|}{(-0.99)}               & \multicolumn{1}{l|}{(-1.02)}                 & \multicolumn{1}{l|}{(-1.00)}                 & (-1.02)     
\end{tabular}
\caption{Change in the reduced chi-square statistics for electron pressure when different features, halo concentration dependence, broken power law (BPL) in mass, and setting $\gamma$ in Eq. (\ref{eqn:10}) as free parameters, are included separately.}
\label{table:2}
\end{table}

\section{Electron density} \label{sec:results_density}

\begin{figure}
\centering
\includegraphics[width=1.0\textwidth]{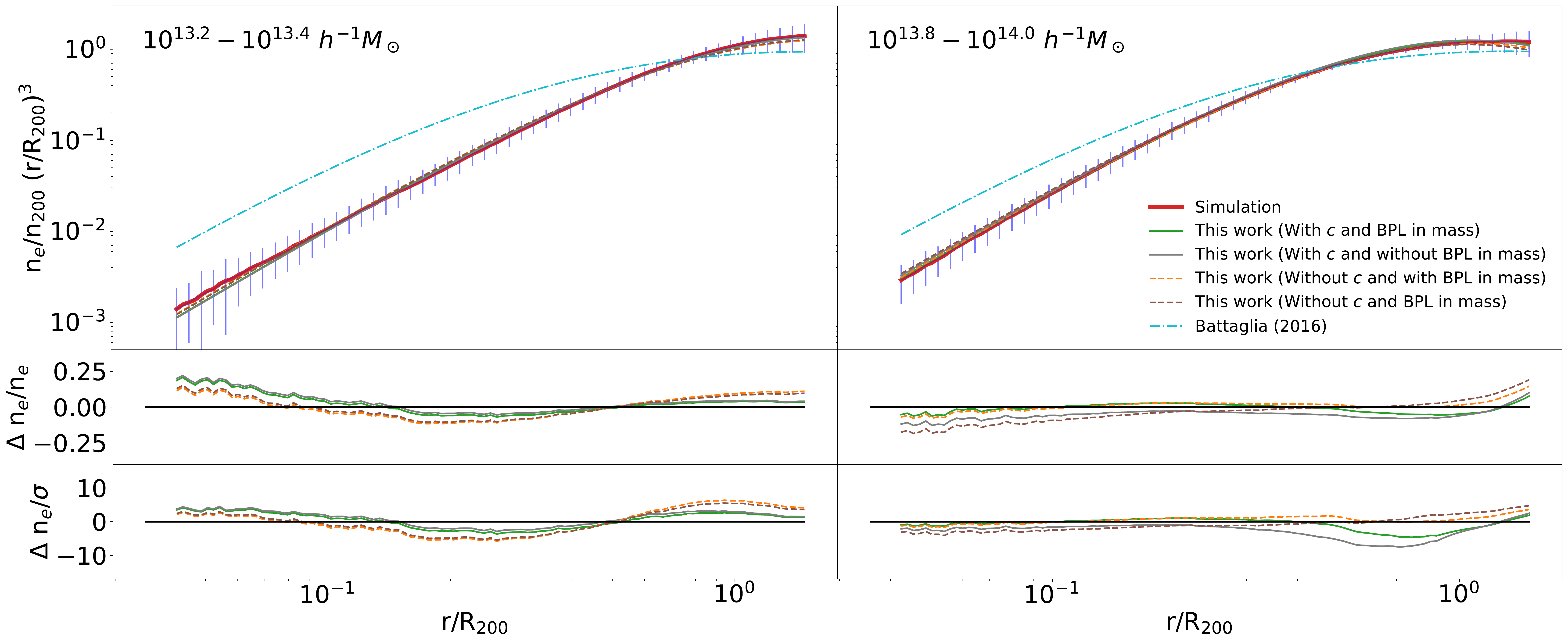}
\caption{Similar to Fig. \ref{fig:2}, but for electron density. The cyan line uses the parameters in \citet{Battaglia:2016xbi}.}
\label{fig:4}
\end{figure}

In an analogous manner, we examine the electron density profiles: The left and right panels of Fig.~\ref{fig:4} show two example profiles of electron density in the lower mass $10^{13.2} - 10^{13.4}$ $h^{-1} M_\odot$ and higher mass $10^{13.8}-10^{14.0}$ $h^{-1} M_\odot$ bin for $z=0.0$ and the lowest concentration in both cases. The cyan line on the plot shows the electron density using the profiles given in \citet{Battaglia:2016xbi}. As in the electron pressure case, we find a generally good fit at large radii and large masses. However, it can be seen that the profile parameters overestimated $n_e/n_{200}$ in the inner region, especially at low masses. We then perform the fitting without $c$ dependence and BPL in mass \citep[the same formalism as in][]{Battaglia:2011cq}, which is indicated by the brown line in the figure. Whilst this updated fit provides a qualitatively better fit, it gives a maximum ratio between the difference between the best-fit model and the data and the standard error of the average electron density profile $\Delta n_e/\sigma = 5.53$ for the lower mass bin.

\begin{figure}
\centering
\plottwo{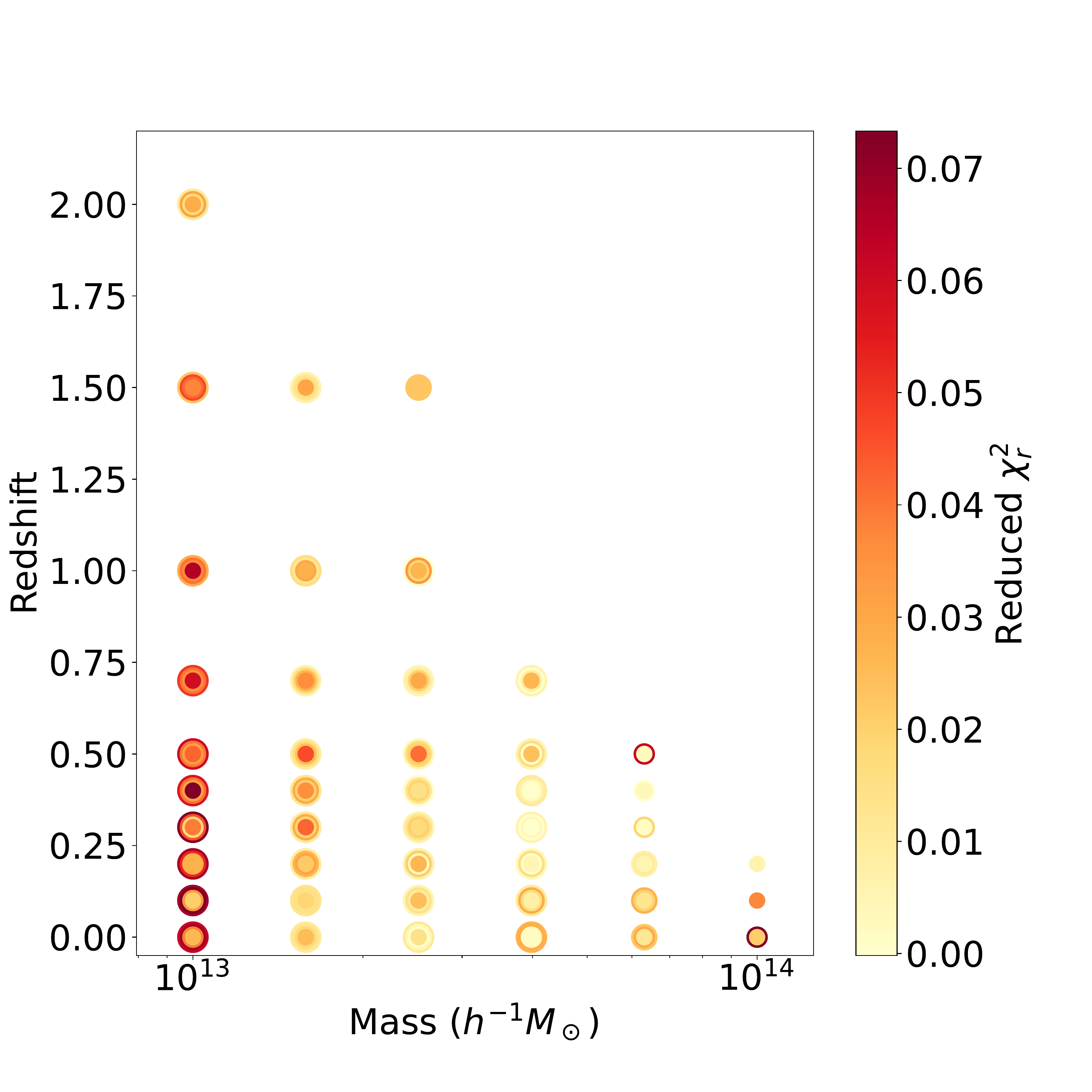}{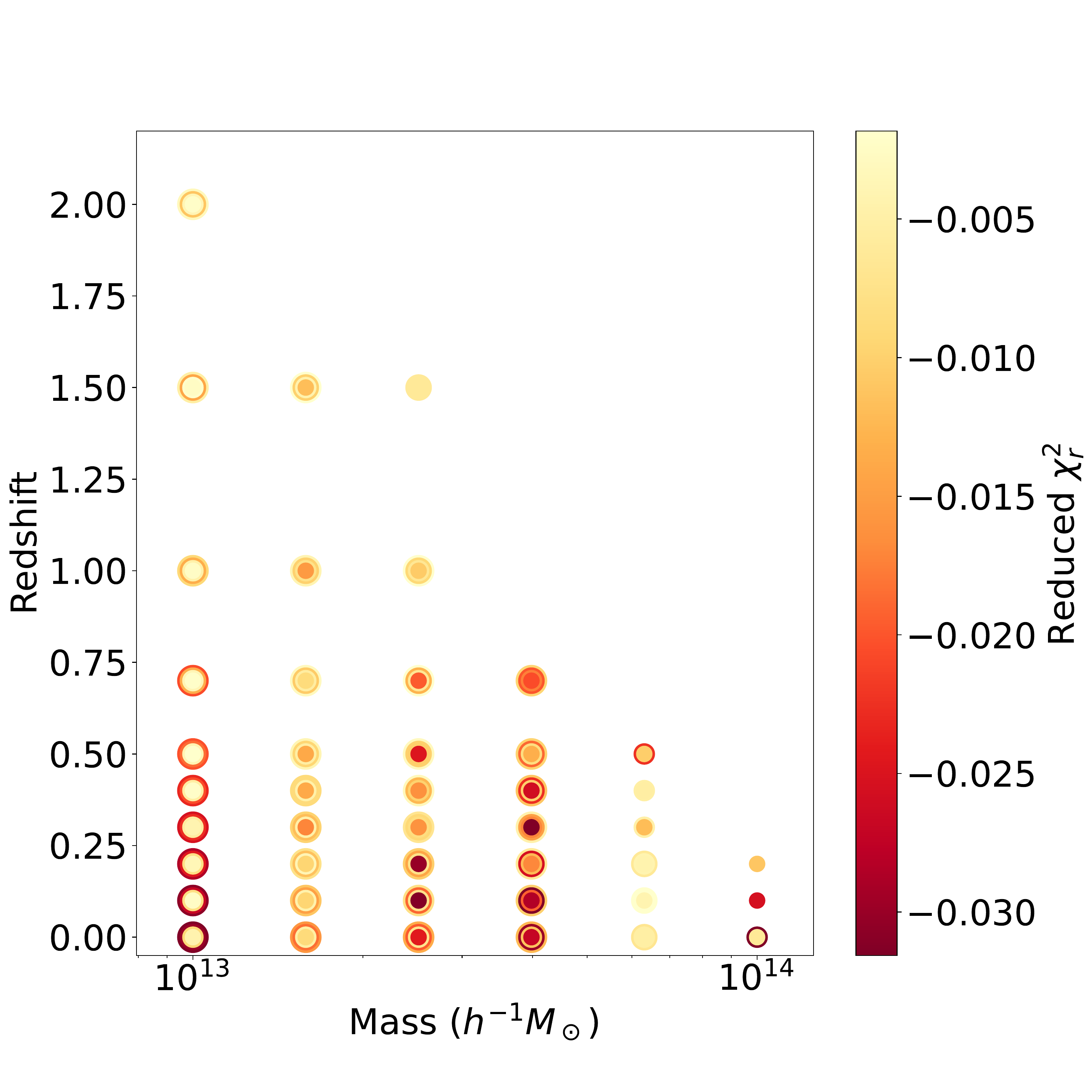}
\caption{Similar to Fig. \ref{fig:3}, but for electron density.}
\label{fig:5}
\end{figure}

\begin{table}
\centering
\begin{tabular}{l|l|l|l|l|l|c}
 & $A_0$ & $\alpha_m$ & $\alpha_m'$ & $\alpha_z$ & $\alpha_c$ & \multicolumn{1}{l}{$\log_{10}M_{\text{cut}}$ ($h^{-1}M_\odot$)} \\ \hline
$n_0$     & $15.7_{-0.3}^{+0.3}$     & $0.87_{-0.01}^{+0.01}$          & --           & $-2.09_{-0.02}^{+0.02}$          & $0.63_{-0.02}^{+0.02}$          & \multirow{3}{*}{$13.75_{-0.01}^{+0.01}$}                    \\
$x'_c$     & $2.2_{-0.1}^{+0.1}$     & $-0.06_{-0.02}^{+0.02}$          & $-1.45_{-0.07}^{+0.07}$           & $-0.74_{-0.05}^{+0.04}$          & $-1.37_{-0.04}^{+0.04}$         &                                       \\
$\beta'$   & $7.5_{-0.2}^{+0.2}$     & $0.24_{-0.02}^{+0.02}$          & $-1.10_{-0.07}^{+0.07}$           & $-0.39_{-0.04}^{+0.04}$         & $-1.11_{-0.04}^{+0.03}$         &                                       \\ 
\end{tabular}
\caption{Best-fit parameters for electron density, Eq. (\ref{eqn:10}) and (\ref{eqn:11}), when halo concentration dependence and broken power law (BPL) in mass in $x'_c$ and $\beta'$ are included.}
\label{table:3}
\end{table}

We then extend the fitting mode to include $c$ dependence and a BPL in mass. In this case, we include a BPL in both $x'_c$ and $\beta'$. This further improves the fit, and the deviations $\Delta n_e/\sigma$ are less than 4.22 for the lower mass bin. We show the best-fit parameters for electron density in Table \ref{table:3} when $c$ dependence and BPL in mass are included. $\alpha_m$, $\alpha_m'$, $\alpha_z$, and $\alpha_c$ all differ from zero significantly, indicating the importance of all these parameters. For example, $n_0$ has $\alpha_z = -2.09^{+0.02}_{-0.02}$, which shows that $\bar{n}_e$ is smaller when redshift is higher. Also, $n_0$ has $\alpha_c = 0.63^{+0.02}_{-0.02}$, showing that a more concentrated halo has a higher $\bar{n}_e$.

We also plot the reduced chi-square statistics in each bin for the electron density in the left panel of Fig. \ref{fig:5}. The same features, where the low mass, low redshift halos have a high chi-square value are observed, and the high mass, high redshift halos have a low chi-square value. Nonetheless, the fact that both electron density and pressure do not have a good fit in the low-mass, low-redshift range might imply that there are intrinsic deficiencies with our one-halo generalized NFW profile, which we will discuss further in the discussion section. In the end, we have an overall chi-square $\chi^2_r = 5.07 \gg 1$ when both $c$ dependence and BPL in mass are included.

Similar to the case of electron pressure, there is a large decrease in the reduced chi-square when we include the two new features. In Table \ref{table:4}, we show the decrease in reduced chi-square when the features are introduced individually. For example, when $c$ dependence and BPL in all parameters are included, the reduced chi-square is reduced by 1.27, compared to 0.56 only with BPL in all parameters. Again, we see that a BPL in all the parameters is not motivated by the data. Hence our best fit model, plotted in Fig. \ref{fig:4} with parameters in Table \ref{table:3}, only uses BPL for $x'_c$ and $\beta'$.

\begin{table}
\centering
\begin{tabular}{lllllllll}
                                            &                           & \multicolumn{7}{c}{Broken power law in}                                                                                                                                                                                                     \\
\multicolumn{1}{l|}{}                       & \multicolumn{1}{l|}{None} & \multicolumn{1}{l|}{$x'_c$} & \multicolumn{1}{l|}{$n_0$} & \multicolumn{1}{l|}{$\beta'$} & \multicolumn{1}{l|}{$x'_c$ and $n_0$} & \multicolumn{1}{l|}{$x'_c$ and $\beta'$} & \multicolumn{1}{l|}{$n_0$ and $\beta'$} & $x'_c$, $n_0$, and $\beta'$ \\ \hline
\multicolumn{1}{l|}{With concentration}     & \multicolumn{1}{l|}{-0.78}    & \multicolumn{1}{l|}{-1.06}     & \multicolumn{1}{l|}{-1.12}     & \multicolumn{1}{l|}{-1.02}       & \multicolumn{1}{l|}{-1.11}               & \multicolumn{1}{l|}{-1.25}                 & \multicolumn{1}{l|}{-1.13}                 & -1.27                         \\
\multicolumn{1}{l|}{(and free $\gamma$ {[}Section 5.2{]})}     & \multicolumn{1}{l|}{(-1.00)} & \multicolumn{1}{l|}{(-1.24)}    & \multicolumn{1}{l|}{(-1.32)}     & \multicolumn{1}{l|}{(-1.20)}     & \multicolumn{1}{l|}{(-1.34)}       & \multicolumn{1}{l|}{(-1.45)}               & \multicolumn{1}{l|}{(-1.34)}                 & (-1.47)    \\ \hline
\multicolumn{1}{l|}{Without concenteration} & \multicolumn{1}{l|}{--}    & \multicolumn{1}{l|}{-0.48}     & \multicolumn{1}{l|}{-0.51}     & \multicolumn{1}{l|}{-0.44}       & \multicolumn{1}{l|}{-0.49}               & \multicolumn{1}{l|}{-0.52}                 & \multicolumn{1}{l|}{-0.48}                 & -0.56     \\
\multicolumn{1}{l|}{(and with free $\gamma$ {[}Section 5.2{]})} & \multicolumn{1}{l|}{(-0.32)}    & \multicolumn{1}{l|}{(-0.66)}     & \multicolumn{1}{l|}{(-0.83)}     & \multicolumn{1}{l|}{(-0.64)}       & \multicolumn{1}{l|}{(-0.82)}               & \multicolumn{1}{l|}{(-0.78)}                 & \multicolumn{1}{l|}{(-0.81)}                 & (-0.82)     
\end{tabular}
\caption{Similar to Table \ref{table:2}, but showing the reduction on reduced chi-square to the fits for electron density when extensions are included in various parameters.}
\label{table:4}
\end{table}

\section{Discussion} \label{sec:discuss}

\subsection{The SIMBA Simulations}

\begin{figure}
\centering
\includegraphics[width=0.6\textwidth]{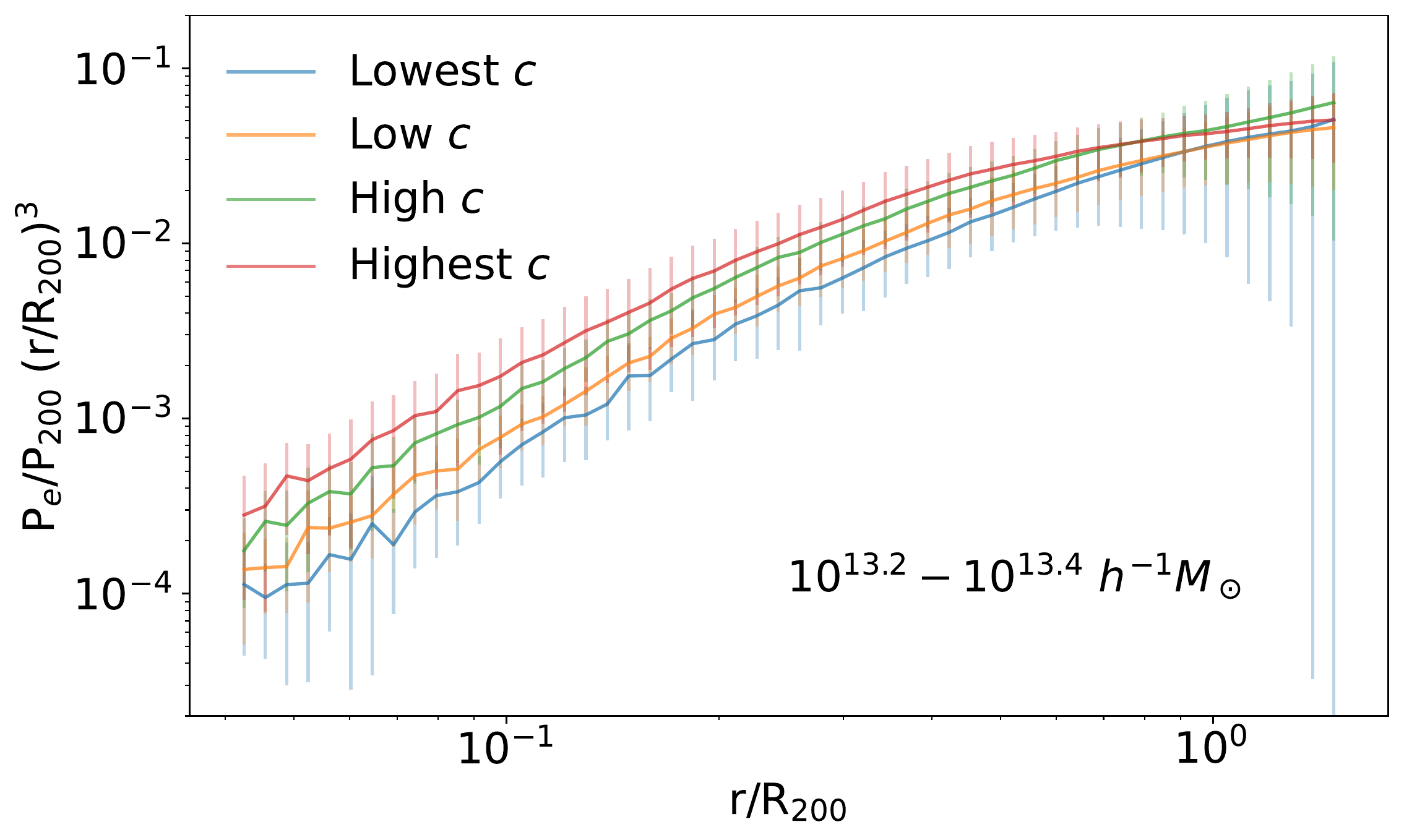}
\caption{Electron pressure profile in the SIMBA simulation for different concentration bins with mass $10^{13.2}-10^{13.4}$ $h^{-1} M_\odot$ and $z = 0.0$.}
\label{fig:6}
\end{figure}

To further examine the robustness of our results, we make use of another independent simulation SIMBA \citep{Dave:2019yyq}. We prepare the data as in the IllustrisTNG simulation. However, due to the smaller volume of the SIMBA simulation compared to the IllustrisTNG simulation, we do not have enough halos to perform the sophisticated fitting using the SIMBA simulation -- we are particularly limited by the small number of massive halos. Therefore, we choose to plot the electron pressure for various concentration bins in the low mass and redshift bin, which has the largest amount of halos. Fig. \ref{fig:6} shows the electron pressure profile in the $10^{13.2}-10^{13.4}$ $h^{-1} M_\odot$ mass bin with 0.0 redshift. An increasing trend in the electron pressure profile is observed when we move to a higher concentration bin, especially in the inner part of the halo, indicating a higher $\bar{P_e}$. This result implies that the $c$ dependence is a general feature that is also observed in simulations different from IllustrisTNG. It is thus expected that further examinations of other state-of-the-art simulations may also find such $c$ dependence. On the other hand, as we only have a few halos for mass higher than $10^{13.6}$ $h^{-1} M_\odot$, we cannot examine whether there is a preferred mass break in the SIMBA simulation.

\subsection{Improving the Fitting}

We found that the low mass, low redshift halos do not fit well to the generalized NFW profile, as indicated by the large reduced $\chi^2_r$ value. However, it is known that lower mass clusters also contribute significantly to the tSZ power spectrum \citep{2010ApJ...725...91B, 2010ApJ...725.1452S}. Methods to reduce the uncertainty in fitting the lower mass halos thus become a relevant issue if we want to achieve the desired accuracy (comparable to the uncertainty given by the halo mass function) in finding the tSZ power spectrum using the fitting formula. There are some possible directions in improving the fitting formula.

\begin{table}
\centering
\begin{tabular}{l|llllll}
 & \multicolumn{1}{l|}{$A_0$} & \multicolumn{1}{l|}{$\alpha_m$} & \multicolumn{1}{l|}{$\alpha'_m$} & \multicolumn{1}{l|}{$\alpha_z$} & \multicolumn{1}{l|}{$\alpha_c$} & $\log_{10}M_{\text{cut}}$ ($h^{-1}M_\odot$)   \\ \hline
$P_0$     & \multicolumn{1}{l|}{$11.9_{-0.2}^{+0.2}$}     & \multicolumn{1}{l|}{$1.07_{-0.01}^{+0.01}$}          & \multicolumn{1}{l|}{$0.75_{-0.01}^{+0.01}$}           & \multicolumn{1}{l|}{$-1.45_{-0.02}^{+0.02}$}          & \multicolumn{1}{l|}{$0.77_{-0.01}^{+0.01}$}          & \multirow{3}{*}{$13.64_{-0.01}^{+0.01}$} \\ 
$x_c$     & \multicolumn{1}{l|}{$0.68_{-0.01}^{+0.01}$}     & \multicolumn{1}{l|}{$-0.24_{-0.01}^{+0.01}$}          & \multicolumn{1}{l|}{--}           & \multicolumn{1}{l|}{$-0.03_{-0.03}^{+0.03}$}          & \multicolumn{1}{l|}{$-0.80_{-0.02}^{+0.02}$}          &                    \\
$\beta$   & \multicolumn{1}{l|}{$5.5_{-0.1}^{+0.1}$}     & \multicolumn{1}{l|}{$0.04_{-0.01}^{+0.01}$}          & \multicolumn{1}{l|}{--}           & \multicolumn{1}{l|}{$0.17_{-0.02}^{+0.02}$}          & \multicolumn{1}{l|}{$-0.39_{-0.02}^{+0.02}$}          &                    \\ \hline
$\gamma$  & \multicolumn{6}{l}{$-0.083_{-0.004}^{+0.004}$}                                                                                                                                                                   \\ 
\end{tabular}
\caption{Best-fit parameters for the electron pressure profiles, Eq. (\ref{eqn:10}) and (\ref{eqn:11}), when including halo concentration dependence, BPL in mass in $P_0$, and setting $\gamma$ to be a free parameter.}
\label{table:5}
\end{table}

\begin{table}
\centering
\begin{tabular}{llllllll}
                             & \multicolumn{7}{c}{$M$ ($h^{-1} M_\odot$)}                                                                                                                                                                                                                                                                \\
\multicolumn{1}{l|}{}        & \multicolumn{1}{l|}{$10^{13.0}-10^{13.2}$}  & \multicolumn{1}{l|}{$10^{13.2}-10^{13.4}$}  & \multicolumn{1}{l|}{$10^{13.4}-10^{13.6}$}  & \multicolumn{1}{l|}{$10^{13.6}-10^{13.8}$} & \multicolumn{1}{l|}{$10^{13.8}-10^{14.0}$} & \multicolumn{1}{l|}{$10^{14.0}-10^{14.2}$} & $10^{14.2}-10^{14.4}$  \\ \hline
\multicolumn{1}{l|}{$P_0$}   & \multicolumn{1}{l|}{$0.85_{-0.02}^{+0.02}$} & \multicolumn{1}{l|}{$1.35_{-0.04}^{+0.04}$} & \multicolumn{1}{l|}{$2.45_{-0.09}^{+0.09}$} & \multicolumn{1}{l|}{$3.9_{-0.2}^{+0.2}$}   & \multicolumn{1}{l|}{$6.5_{-0.4}^{+0.5}$}   & \multicolumn{1}{l|}{$8.4_{-0.8}^{+0.8}$}   & $21_{-4}^{+5}$         \\
\multicolumn{1}{l|}{$x_c$}   & \multicolumn{1}{l|}{$2.3_{-0.1}^{+0.1}$}    & \multicolumn{1}{l|}{$2.5_{-0.1}^{+0.2}$}    & \multicolumn{1}{l|}{$2.1_{-0.1}^{+0.1}$}    & \multicolumn{1}{l|}{$2.0_{-0.1}^{+0.1}$}   & \multicolumn{1}{l|}{$1.4_{-0.1}^{+0.1}$}   & \multicolumn{1}{l|}{$1.2_{-0.1}^{+0.2}$}   & $0.39_{-0.05}^{+0.06}$ \\
\multicolumn{1}{l|}{$\beta$} & \multicolumn{1}{l|}{$7.7_{-0.3}^{+0.3}$}    & \multicolumn{1}{l|}{$9.3_{-0.4}^{+0.4}$}    & \multicolumn{1}{l|}{$9.5_{-0.4}^{+0.5}$}    & \multicolumn{1}{l|}{$10.1_{-0.5}^{+0.5}$}  & \multicolumn{1}{l|}{$8.5_{-0.4}^{+0.5}$}   & \multicolumn{1}{l|}{$8.4_{-0.6}^{+0.8}$}   & $4.3_{-0.2}^{+0.2}$   
\end{tabular}
\caption{Best-fit parameters of the electron pressure profiles, Eq. (\ref{eqn:10}), when fitting each mass bin independently. This demonstrates the complex mass evolution of the parameters.}
\label{table:6}
\end{table}

Throughout the calculation, we follow previous literature by fixing the inner slope parameter $\gamma$, such that $\gamma = -0.3$ for the generalized NFW profile, Eq. (\ref{eqn:10}). However, we observe in Fig. \ref{fig:2} and \ref{fig:4} that the inner region of the profiles shows the largest percentage difference with the simulation data when no $c$ dependence or BPL in mass is imposed. Thus, this motivates us to examine the case where $\gamma$ is set to be a free parameter. For the case with $\gamma$ as a free parameter, $c$ dependence, and BPL in mass for $P_0$, the best-fit electron pressure parameters are shown in Table \ref{table:5}. Note that the best-fit $\gamma = -0.08$ differs from the $\gamma = -0.3$ previously assumed in the literature. When we have $\gamma$ to be a free parameter, $c$ dependence, and BPL in mass for $P_0$, the reduced chi-square is decreased by 2.90 compared to no features imposed. Still, the reduced chi-square value $\chi^2_r = 2.4 \gg 1$ indicates that we do not have a good fit even when all features are included.

Also, we have assumed a simple power-law dependence with mass, redshift, and concentration for parameters $P_0$, $n_0$, $x_c$, $x'_c$, $\beta$, and $\beta'$ so far. We can test this assumption by fitting the electron pressure profile, Eq. (\ref{eqn:10}), at $z = 0$ independently for each mass bin and examine the mass dependence of $P_0$, $x_c$, and $\beta$. Table \ref{table:6} shows the results of fitting $P_0$, $x_c$, and $\beta$ in each mass bin. Note that $\beta$ first goes up until $10^{13.6} - 10^{13.8}$ $h^{-1} M_\odot$ and then goes down for higher mass. A similar uptrend and downtrend can be observed for $x_c$. This uptrend and downtrend impose a problem to our assumption, as the power-law function is monotonic, and the function cannot capture this behaviour. It is possible that using a more complex normalization of the profiles could help. For example, \citet{Lau_2015} found that the inner regions are more self-similar when normalized by $r_{200c}$, whereas the outskirts are more self-similar when normalized by $r_{200m}$. Such results motivate an exploration of functional forms beyond the generalized NFW profile, and one way to explore the function space would be using symbolic regression. Instead of fitting parameters in one particular formula, the symbolic regression algorithm will help us to look for the best function that describes the electron pressure and density. With the help of advancing computational power and algorithms for symbolic regression, we can overcome the limitations of the generalized NFW profile.

It is known that the circumgalactic medium is affected significantly by the AGN feedback strength, and the gas density deviates from the generalized NFW profile. Therefore, examining the effect of different feedback mechanisms on the fitting will be useful for investigating the bad fit in the low mass, low redshift halos. However, there is only one feedback model in the IllutrisTNG simulation because running a full-sized simulation with different feedback models will take an undesirable time. Some work has been done highlighting the variance across different hydrodynamic simulations \citep{Kim:2021gtf}, but an exhaustive search on cosmological volumes remains a challenge. On the other hand, the recent development of the Cosmology and Astrophysics with MachinE Learning Simulations (CAMELS) that includes extra parameters for the stellar and AGN feedback could be helpful for us to explore the effects of feedback on the fitting further \citep{Villaescusa-Navarro:2020rxg}, though the small simulation box size makes it challenging to study the properties of the most massive groups and clusters.

Another assumption worth revisiting is the assumption of sphericity. Asphericity and substructure do contribute to the tSZ power spectrum at multipole moment $l \sim 2000-8000$ \citep{Battaglia:2011cq}, increasing the tSZ power spectrum by 10-20\%. Through spherical averaging, we have transferred the triaxiality of the cluster profiles into an additional source of noise. Whilst a recent study of the IllustrisTNG simulations \citep{Thiele_2022} suggests that asphericity does not play a major role in the profiles of galaxy clusters, a systematic investigation of this is needed. In addition, we have assumed a one-halo generalized NFW profile to describe the halo. Additional terms such as the two-halo term, which accounts for the effects from neighbouring halos and gases that are not virialized, are important to describe the radial profile when $r > R_{200}$. Thus, the two-halo term could be important in the fitting, as suggested by various research articles \citep{Vikram:2016dpo, Moser:2021llm, Hill:2017tua}. Note that probing the 1/2 halo boundary is challenging in these simulations - both simulations make the strong assumption that the temperature of the electrons and gas are equal, which is not the case \citep{Rudd_2009,Avestruz_2015}, and this assumption can lead to a large underestimate of the pressure in the cluster outskirts. Whilst we chose our radial cut off to minimize our sensitivity to the two-halo terms, the inclusion of a two-halo term would give extra flexibility to the fitting formula. A thorough investigation of the importance of these effects is left to future work.

\section{Conclusion} \label{sec:conclusion}

In conclusion, we studied the profiles of the electron pressure and density in the IllustrisTNG and SIMBA numerical cosmology simulations. We provide an updated version of the commonly used \citet{Battaglia:2011cq} fitting formula for these profiles that provides a more accurate description of the lower mass halos. 

Building on this, we investigated the impact of extended parameterizations. Recent observations \citep{Pandey_2021} have suggested that cluster $y$-profiles exhibit deviations from a simple power law in mass, and our results find that this is replicated in our simulations - a BPL provides a better fit to our cluster profiles. Beyond power-law behaviour can naturally arise in analytical feedback models \citep[e.g.][]{Shaw_2010}, and it will be interesting to explore how constraints on this break can be used to enhance our understanding of feedback processes. Interestingly, we find that both the pressure profiles and electron density profiles prefer similar masses ($4 - 6 \times 10^{13} M_\odot$) for the break in the power law. This could indicate the boundary between two qualitatively different regimes of the impact of the feedback process on the gas. Note that our mass break is lower than that used in \citet{Pandey_2021}. More significantly, the introduction of $c$ dependence into the fitting formula provides vastly improved fits. It suggests that we may not achieve the desired accuracy in the tSZ power spectrum without considering this important factor.

Observations of cluster electron pressure and density profiles can be a useful tool to understand the astrophysical process. However, to interpret the results, it is crucial to understand the covariance matrix. Our results find quantitatively different behaviour in the covariance matrix for high and low mass halos, with significantly stronger radial correlations in massive halos. This behaviour, driven by the difference in AGN and supernova feedback, is to be accounted for in observational studies to avoid biased inferences. We also see a similar effect in the SIMBA simulations, hinting that this feature may be a generic feature. 

By investigating the measures of the goodness of fit, we find that, despite our extensions, our best fit model does not capture the electron pressure and density profiles for low mass and low redshift halos in IllustrisTNG. It suggests that the generalized NFW profile is not suitable in this regime, and it highly motivates further study into the methods that reduce the error. One such direction would be using symbolic regression to identify the best profile representing the electron pressure and density.

A comparison between the IllustrisTNG and SIMBA simulations (Fig. \ref{fig:2} and Fig. \ref{fig:6}) highlights how strongly the details of baryonic processes impact observed cluster profiles. Understanding how our modelling uncertainty can be included in this framework represents the biggest challenge and requires simulations that cover a wide range of plausible subgrid models, such as those in the CAMELS suite of simulations \citep{Villaescusa-Navarro:2020rxg}.

Despite these caveats, our fitting formula has a range of uses, from combining with dark matter simulations to generate simulated SZ observations \citep[e.g][]{Stein_2020} to providing more realistic halo model calculations and modelling of observations \citep{Amodeo:2020mmu}. More broadly, the necessity of including $c$ dependence, non-power law dependence, and mass-dependent radial correlations are applicable to more general methods of modelling SZ observations.

\appendix

\begin{table}[h]
\centering
\begin{tabular}{c|c|c|c|c|c}
 & $A_0$ & $\alpha_m$ & $\alpha_m'$ & $\alpha_z$ & \multicolumn{1}{c}{$\log_{10}M_{\text{cut}}$ ($h^{-1}M_\odot$)} \\ \hline
$P_0$     & $2.8_{-0.1}^{+0.1}$     & $1.08_{-0.01}^{+0.01}$          & $0.80_{-0.01}^{+0.01}$           & $-1.89_{-0.02}^{+0.02}$          &  \multirow{3}{*}{$13.60_{-0.01}^{+0.01}$}                    \\ 
$x_c$     & $2.10_{-0.04}^{+0.04}$     & $-0.35_{-0.02}^{+0.02}$          & --           & $0.53_{-0.04}^{+0.04}$          &                                       \\ 
$\beta$   & $9.4_{-0.1}^{+0.1}$     & $-0.06_{-0.01}^{+0.01}$          & --           & $0.60_{-0.03}^{+0.04}$          &                                       \\ 
\end{tabular}
\caption{Best-fit parameters for the electron pressure profiles, Eq. (\ref{eqn:10}) and (\ref{eqn:11}), when broken power law (BPL) in mass in $P_0$ is included.}
\label{table:7}
\end{table}

\begin{table}[h]
\centering
\begin{tabular}{l|l|l|l|l|c}
 & $A_0$ & $\alpha_m$ & $\alpha_m'$ & $\alpha_z$ &  \multicolumn{1}{l}{$\log_{10}M_{\text{cut}}$ ($h^{-1}M_\odot$)} \\ \hline
$n_0$     & $6.8_{-0.1}^{+0.1}$     & $0.68_{-0.01}^{+0.01}$          & --           & $-2.11_{-0.02}^{+0.02}$          & \multirow{3}{*}{$13.61_{-0.01}^{+0.01}$}                    \\
$x'_c$     & $7.9_{-0.2}^{+0.2}$     & $0.47_{-0.03}^{+0.03}$          & $-0.45_{-0.03}^{+0.03}$           & $-0.67_{-0.04}^{+0.04}$                &                                       \\
$\beta'$   & $19.5_{-0.5}^{+0.5}$     & $0.70_{-0.03}^{+0.03}$          & $-0.18_{-0.03}^{+0.03}$           & $-0.31_{-0.03}^{+0.03}$            &                                       \\ 
\end{tabular}
\caption{Best-fit parameters for the electron density, Eq. (\ref{eqn:10}) and (\ref{eqn:11}), when broken power law (BPL) in mass in $x'_c$ and $\beta'$ is included.}
\label{table:8}
\end{table}


\newpage
\bibliography{sample631}{}
\bibliographystyle{aasjournal}



\end{document}